# Increasing Opportunistic Gain in Small Cells Through Energy-Aware User Cooperation

Qing Wang, *Student Member, IEEE,* Balaji Rengarajan, *Member, IEEE,* and Joerg Widmer, *Senior Member, IEEE*

***Abstract*—To meet the increasing demand for wireless capacity, future networks are likely to consist of dense layouts of small cells. The number of users in each cell is thus reduced which results in diminished gains from opportunistic scheduling, particularly under dynamic traffic loads. We propose an user-initiated *base station (BS)-transparent* traffic spreading approach that leverages user-user communication to increase BS scheduling flexibility. The proposed scheme can increase opportunistic gain and improve user performance. For a specified tradeoff between performance and power expenditure, we characterize the optimal policy by modeling the system as a Markov decision process and also present a heuristic algorithm that yields significant performance gains. Our simulations show that, in the performance-centric case, average file transfer delays are lowered by up to 20% even in homogeneous scenarios, and up to 50% with heterogeneous users. Further, we show that the bulk of the performance improvement can be achieved with a small increase in power expenditure, e.g., in an energy-sensitive case, up to 78% of the performance improvement can be typically achieved at only 20% of the power expenditure of the performance-centric case.**

## I. Introduction

Opportunistic scheduling [2, 3] was proposed for multiuser wireless communication networks to exploit fluctuating channel conditions, aiming to improve performance. In cellular networks, opportunistic schedulers use knowledge of the channels between Base Station (BS) and users to schedule those with favorable channel states, thus improving overall throughput.

The performance of opportunistic scheduling algorithms has been commonly investigated under the assumption of a static user population with infinitely backlogged queues [3], i.e., the BS always has data to transmit to each user. However, a more realistic setting is one with a time-varying user population and stochastic traffic loads. In such a setting, the performance of opportunistic scheduling algorithms can be very different [4, 5]. The impact of a time-varying user population is small in large cells since the BS may always have a large number of users to choose from for scheduling purposes. However, as cell sizes in future wireless networks shrink in response to increasing demands for capacity [6–8], the average number of users served by a BS will decrease and the burstiness will increase. Since opportunistic gain scales as a concave function of the user population [9], presently used scheduling algorithms are prone to losing effectiveness in small cells with dynamic traffic load.

In this paper, we propose an alternate user-initiated *BS-transparent* (i.e., without changes at the BS) algorithm. We focus on the downlink case which accounts for most of the traffic in a cellular network [10], and on http live streaming or best-effort traffic (e.g., web browsing), where mobile devices request files (chunks of content) which are then sent to users after some fetching and queueing delay. To spread traffic, we leverage the multiple radio interfaces (e.g. 3G, WiFi) available in most smartphones. The algorithm keeps tracking each user' backlogs at the BS and balances traffic requests across users, aiming to maximize the BS's long-term scheduling options and hence improve the delay performance. This improved performance on delay implicitly reduces the power consumption of cellular transmission [11].

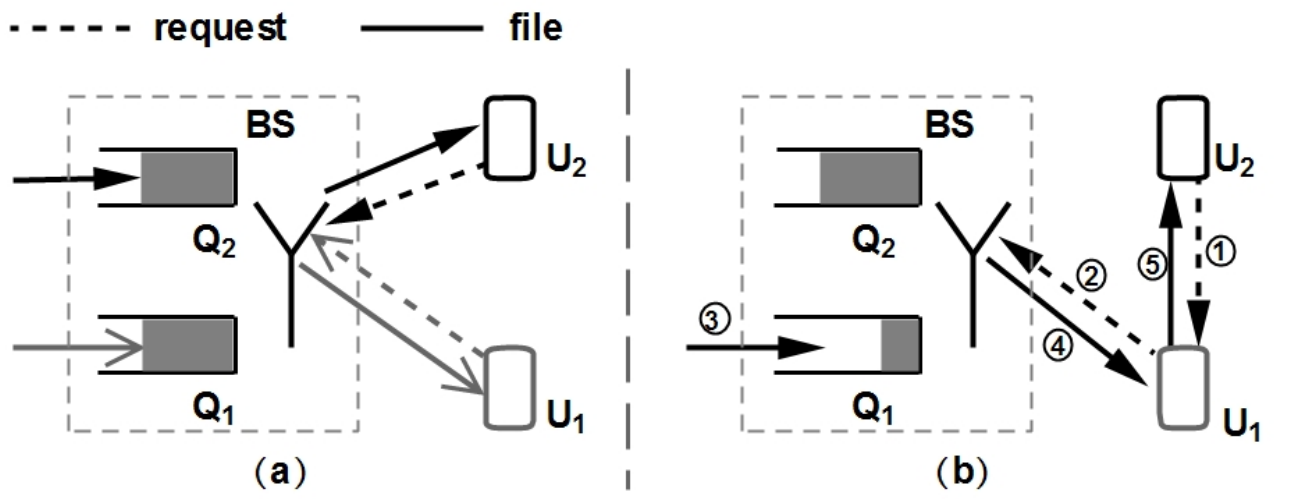

Fig. 1: An example of traffic spreading: (a) No traffic spreading; (b) Traffic spreading from $U_2$ to $U_1$.

Our proposed algorithm includes a dispatcher that resides on each mobile device. To illustrate the traffic spreading, let us consider an example shown in Fig. 1. We depict a scenario with two users, $U_1$ and $U_2$ being served by the BS. The queues, $Q_1$ and $Q_2$, depict the number of files at users' BS queues. In this scenario, the users perceive similar channel statistics and we consider below the case where they generate similar traffic loads to illustrate the spreading mechanism. In Fig. 1(a), since the queues at the BS are balanced, the dispatchers of the users would ideally detect that traffic spreading is not beneficial. Thus users send their new requests to the BS directly. In Fig. 1(b), there are many more files in $Q_2$ than in $Q_1$ which is nearly empty. Under this case, the dispatcher of $U_2$ would ideally detect that traffic spreading is beneficial since in the near term the risk is high that the BS has no files to send to $U_1$ and thus losses opportunistic gain. Thus, when a new request is generated by $U_2$, it forwards the request to $U_1$, who will send it to the BS. When $U_1$ receives the corresponding file from the BS, it forwards the file to $U_2$ through the user-to-user

Q. Wang (corresponding author) is with the IMDEA Networks Institute and the University Carlos III of Madrid, Spain. E-mail: qing.wang@imdea.org

B. Rengarajan is with the Accelera Mobile Broadband, CA, USA. E-mail: balaji.rengarajan@gmail.com

J. Widmer is with the IMDEA Networks Institute, Madrid, Spain. E-mail: joerg.widmer@imdea.org

link.

The decision made by dispatchers is based on the channel statistics of all the users and their backlogs at the BS. Note that dispatchers do not exploit current channel conditions, and can not predict when a new request will be served or the instantaneous channel conditions at that time. Each user keeps track of the number of files at the BS, and shares this information periodically with other users. Users also measure their perceived channel statistics and exchange them with other users. Moreover, users are aware of, or can easily infer the BS scheduling policy and can track the destination of files received from the BS. Note that our proposed algorithm also applies to the scenarios where some users do not have their own file requests. These users will act as pure relaying nodes.

The proposed algorithm incurs additional power expenditure due to forwarding requests and files among users. Mobile devices with scarce energy resources necessitate careful power management because excessive traffic spreading can result in unacceptably high penalties in terms of power expenditure. Thus, the degree of spreading has to be carefully chosen, and the tradeoff between performance improvement and additional power expenditure must be taken into account. In this paper, we develop an energy-aware traffic spreading policy that can optimize the degrees of spreading based on desired performance-energy tradeoff. We summarize our main contributions as follows:

1) We propose a novel traffic spreading policy to increase opportunistic gains by energy-aware user cooperation.
2) We formulate the problem of determining the optimal spreading policy under a specified tradeoff between performance and energy as a Markov decision problem, and study properties of the corresponding optimal policy in a two-user scenario.
3) We propose a heuristic algorithm to reduce the computational complexity in large systems by aggregating users and using the two-user solution as a building block.
4) Based on realistic Rayleigh fading channels, we provide simulation results demonstrating that under our proposed policy: $i$) average file transfer delays be reduced by up to $20\%$ even in homogeneous scenarios using the proposed methodology; $ii$) significant gains (up to 78%) are typically achieved at only $20\%$ of the power expenditure of the performance-centric case; $iii$) the delay performance can be improve greatly (by up to 50%) in scenarios where the overall system performance is poor when some users locate far from the BS.
5) We also extend our proposed traffic spreading algorithm to large cells. Evaluation results show that in a large cell, the delay performance can be improved by up to $50\%$, and up to $73\%$ of the gain can be achieved at only $18\%$ of the maximal additional power expenditure.

Compared to our previous work [1], we analyze the algorithm in more details, prove its properties, run more simulations, extend the proposed traffic spreading policy to large cells and evaluate it. The rest of this paper is organized as follows: related work is summarized in Sec. II, followed by system model and dynamic programming formulation in Sec. III and IV respectively. The properties of the optimal traffic spreading policy are described in Sec. V, and a tractable heuristic for large systems is developed in Sec. VI. Simulation results and evaluation of the heuristic are presented in Sec. VII. Finally, we present our conclusions and future work in Sec. IX.

## II. Related Work

Many opportunistic scheduling algorithms taking into account users' backlogs during BS scheduling have been proposed for systems with dynamic traffic. Authors in [12–14] propose BS schedulers that try to maximize opportunistic gain as well as balance users' backlogs. Among these, [12, 13] propose the throughput-optimal MaxWeight and Exponential rules, respectively, and [14] proposes a policy named log rule to improve delay performance. All these policies react to imbalance in users' queues, sacrificing opportunistic gain in order to balance the queues. They also necessitate changes at the BS. In contrast, the policy proposed in our work is able to balance users' backlogs without sacrificing opportunistic gain, by opportunistically exploiting the BS-user and user-user channels. Moreover, our policy is *BS-transparent* and thus does not require any changes at the BS.

Some approaches are proposed to exploit both the BS-user and user-user channels, e.g., *opportunistic relaying* [15–17] and *device-to-device communication* [18–20]. Among these, [15] proposes the idea of opportunistic relaying as well as an approach that chooses the best relay maximizing the minimal quality of BS-relay and relay-user channels. In [16, 17], mobile users themselves are used as relays, instead of particular relay nodes. In [18–20], users are divided into clusters and in each cluster, a cluster header is responsible to communicate with the BS and forwards traffic to other users. Compared to ours, [15, 16, 18–20] assume users have infinitely backlogged queues. While [17] considers stochastic traffic loads as ours, none of [15–20] have investigated the delay-energy tradeoff.

As we describe in Sec. III, we formulate the problem of determining the optimal traffic spreading policy as a dispatching problem. Here, we discuss some related work on this topic. A *dispatching system* typically consists of a dispatcher and several servers. The role of the dispatcher is to route new jobs to a server based on dispatching policies. The dispatching problem has received a lot of attention since the landmark work in [21]. The author considers a homogeneous model with Poisson arrivals and exponentially distributed job size, and show that when the queue lengths of the servers (number of jobs) are known, Join the Shortest Queue (JSQ) minimizes the average waiting time in the queues. When queue lengths are unavailable, [22] shows that Round Robin is optimal. In contrast to above papers, our model only has one shared server whose service rate is affected by the dispatching policy. The model in [23] includes the case of a shared server and is the closest to ours. However, this paper like the others makes the assumption that the service rate is constant and does not depend on the instantaneous queue states. In our work, the service rate depends on the channel states as well as queue states, making the problem more complex. The emphasis in all the above papers is on performance, whereas in our case we additionally consider the implications of the dispatching decisions on energy cost.

## III. System Model

We model the system in continuous time with $N$ users attached to a single BS, where the set of users is denoted by $\mathcal{I} = \{1, 2, ..., N\}$. The arrival requests of users are modeled as Poisson processes with mean arrival rate vector $\boldsymbol{\lambda} = \{\lambda_1, \lambda_2, ..., \lambda_N\}$, and are assumed to be independent across users. The requested file sizes of users are exponentially distributed, with mean file size vector $\boldsymbol{\theta} = \{\theta_1, \theta_2, ..., \theta_N\}$.

**Channel model:** The wireless channel is assumed to be time-varying and the channel instance between the BS and users can take values from the set $\mathcal{S} = \{1, 2, \cdots, K\}$. We denote by $\boldsymbol{C}(t) = \{C_i(t), i \in \mathcal{I}, C_i(t) \in \mathcal{S}\}$, the vector of users' current channel states at time $t$ and by $\boldsymbol{\mathcal{C}}$, the set of all the possible channel state vectors. Further, we assume the channels perceived by different users are independent. We denote by $p_i^k, k \in \mathcal{S}$, the probability that user $i$ perceives channel $k$ at any time. Each user $i$ shares its channel probability vector $\boldsymbol{p}_i = \{p_i^1, p_i^2, ..., p_i^K\}$ with other users. We assume all users are within the transmission range of each other, which is expected to be the case in picocell/femtocell [24, 25] scenarios (*the system model is also extended to large cells in Sec. VIII, where not all users can communicate with each other*). We define a non-negative value $R_i^k$ for each channel state $k \in \mathcal{S}$, which denotes the data rate (rate supported by the channel) in bits/second of user $i$.

Our system model consists of three main components, i.e., the BS scheduler, the queues at the BS, and a dispatcher that models the joint behavior of all the mobile devices. The BS maintains a separate queue corresponding to each user, and we denote by $\boldsymbol{Q}(t) \equiv (Q_i(t), i \in \mathcal{I}) \in \mathbb{Z}_+^N$, the number of files waiting to be sent by the BS to each user at time $t$, i.e., the number of pending requested files. We follow the convention that random variables are denoted by capital letters (i.e., $\boldsymbol{Q}(\cdot)$ and $\boldsymbol{C}(\cdot)$), while the possible values are denoted by the corresponding small letters (i.e., $\boldsymbol{q}$ and $\boldsymbol{c}$).

**Scheduling policy:** We model the BS scheduling policy through $\xi_i(\boldsymbol{q}, \boldsymbol{c})$, denoting the probability that user $i$ is selected to be served by the scheduler, conditional on the queues being in state $\boldsymbol{q}$, and channel vector being $\boldsymbol{c}$. The average queue state-dependent service rate of user $i$ is denoted as $\mu_i(\boldsymbol{q})$. We consider two channel-aware scheduling policies: a queue-unaware policy where $\xi_i(\boldsymbol{q}, \boldsymbol{c})$ only depends on the set of non-zero elements in $\boldsymbol{q}$, and a queue-aware policy that have a stronger dependence on $\boldsymbol{q}$:

*1) Queue-unaware, greedy scheduling policy:* Queue-unaware means the scheduler is unaware of the queue length, but knows whether a queue is empty or not. At any time $t$, a greedy scheduler chooses a non-empty queue $i$ to serve if user $i$ has the largest instantaneous data rate.

*2) Queue-aware, log rule scheduling policy [14]:* Queue-aware means the scheduler is aware of the queue length. At time $t$, a log rule scheduler makes decisions based on current channel state and the logarithm of queue length, i.e., choosing user $i$ that satisfies

$$i \in \arg\min_{j \in \mathcal{I}} \frac{R_j^{c_j(t)}}{\sum_{k \in \mathcal{S}} p_j^k R_j^k} \log(b + a_j Q_j(t)) \ ,$$

where $b$ and $a_j$ are constants.

**Dispatching policy:** The dispatching policy used across all users when the local communication among users is not congested is defined through the probability matrix $\boldsymbol{\sigma}(\boldsymbol{q})$ as:

$$\boldsymbol{\sigma}(\boldsymbol{q}) = \begin{bmatrix} \sigma_1^1(\boldsymbol{q}) & \cdots & \sigma_1^N(\boldsymbol{q}) \\ \cdots & \cdots & \cdots \\ \sigma_N^1(\boldsymbol{q}) & \cdots & \sigma_N^N(\boldsymbol{q}) \end{bmatrix}, \tag{1}$$

where $\sigma_j^i(\boldsymbol{q})$ denotes the probability of dispatching user $j$'s request to user $i$, conditional on the queues being in state $\boldsymbol{q}$. If the local communication is congested, then users will never dispatch their requests to other users. The set of all the possible dispatching policies is defined as

$$\mathcal{A} \equiv \{\boldsymbol{\sigma}(\boldsymbol{q}) : \sum_{i \in \mathcal{I}} \sigma_j^i(\boldsymbol{q}) = 1, \ 0 \leq \sigma_j^i(\boldsymbol{q}) \leq 1, \ j \in \mathcal{I}\} \ . \tag{2}$$

The rate at which files arrive to user $i$'s queue at the BS is denoted by $\lambda_i'(\boldsymbol{q}, \boldsymbol{\sigma}(\boldsymbol{q}))$. This rate corresponds to the rate of requests sent by user $i$ to the BS (including forwarded requests from other users), as shown in Fig. 2 (a). The insight of the dispatching policy is illustrated in Fig. 2 (b), where we depict a two-user scenario and the axes represent users' BS queues. The $\mu_i'$ and $\mu_i$ are user $i$'s service rate when its queue is and is not empty, respectively. It is straightforward that $\mu_i' \geq \mu_i, i \in \mathcal{I}$. As opportunistic gain scales with the number of users [9], we have $\sum_{i=1}^2 \mu_i \geq \mu_j', \forall j \in \{1, 2\}$, namely, opportunistic gain decreases when queue state reaches the axes. Our proposed dispatching policy dynamically controls the direction of vector $\boldsymbol{\lambda}'$ to keep as many queues being non-empty as possible, which helps scheduling policies to increase the opportunistic gain.

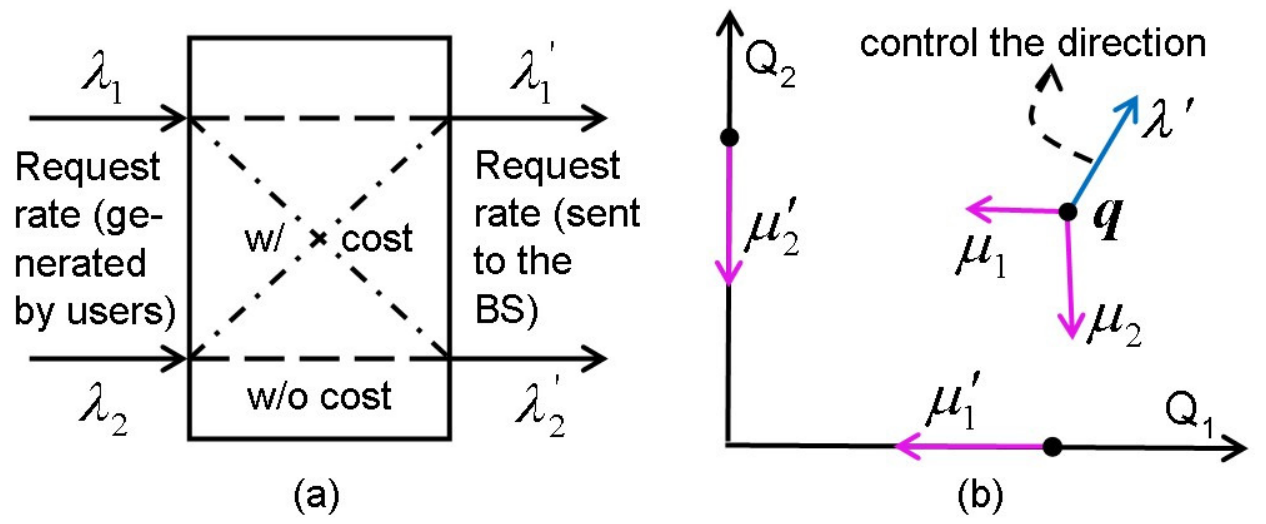

Fig. 2: Proposed dispatching policy: (a) the dispatcher; (b) the dispatching algorithm controls the arrivals to users' BS queues.

The proposed policy also applies to multicast. The difference in multicast is all the users subscribing the same multicast channel have identical arrival requests. Instead of sending the same request by all users, they can choose the one with the shortest BS queue to send a file request. This chosen user then forwards the corresponding file to others after receiving it.

**Performance metrics:** The metrics we use are average file transfer delay and *the re-routing cost*, i.e., additional power expenditure induced by traffic spreading. We assume there is no queueing delay for the user-to-user transfer, but each transfer certainly adds additional forwarding delay and power expenditure. We model the additional delay and power expenditure incurred for a single file of user $j$ as a function of its mean file size $\theta_j$, i.e., $\eta_j^i(\theta_j)$ and $\phi_j^i(\theta_j)$, and define $\eta_j^j(\theta_j)=0$, $\phi_j^j(\theta_j)=0$. Note that a very high $\phi_j^i(\theta_j)$ can be used to model the case where the large distance between users makes the communication between them infeasible. Moreover, we denote by $\bar{\eta}_j^i$ and $\bar{\phi}_j^i$, $i,j\in\mathcal{I}$, the average additional forwarding delay and power expenditure incurred for re-routing files of user $j$, respectively. The objective function we seek to minimize is

$$\bar{D}+\sum_{j\in\mathcal{I}}\sum_{i\in\mathcal{I}}\bar{\eta}_j^i+\sum_{j\in\mathcal{I}}\sum_{i\in\mathcal{I}}w_j^i\cdot\bar{\phi}_j^i\ , \tag{3}$$

where $\bar{D}$ is the average BS-user delay and $\boldsymbol{w}=\{w_j^i,i,j\in\mathcal{I}\}$ is the $N\times N$ weight matrix associated with the power expenditure of users that determines the tradeoff between delay and power expenditure.

## IV. Dynamic Programming Formulation

Consider the process $(\boldsymbol{Q}(t),t\geq 0)$ initiated in state $\boldsymbol{Q}(0)$ and evolving under a dispatching policy $\boldsymbol{\sigma}$ and a scheduling policy $\boldsymbol{\xi}$. We define the vector $\boldsymbol{\mu}(\boldsymbol{q})\equiv(\mu_i(\boldsymbol{q}),i\in\mathcal{I})$ as the average service rate of users, if the queue state is $\boldsymbol{q}$. Clearly, $\boldsymbol{\mu}(\boldsymbol{q})$ depends on the scheduling policy used at the BS. For a given scheduling policy, the average service rate $\mu_i(\boldsymbol{q})$ is given as

$$\mu_i(\boldsymbol{q})=\sum_{\boldsymbol{c}\in\boldsymbol{\mathcal{A}}}\left(\prod_{j\in\mathcal{I}}p_j^{c_j}\right)\xi_i(\boldsymbol{q},\boldsymbol{c})\cdot R_i^{c_i}\ . \tag{4}$$

We assume over an epoch, each queue $i\in\mathcal{I}$ is served at constant service rate $\mu_i(\boldsymbol{q})$. Since the channel varies much faster than the queue dynamics, we use the service rate averaged across channel fluctuations at each queue state. A rigorous justification of the service rate with consideration of packet or file dynamics can be found in [26]. If the process is in state $\boldsymbol{q}$ and under the dispatching policy $\boldsymbol{\sigma}(\boldsymbol{q})$, the file arrival rate to $Q_i$ is given by

$$\lambda_i'(\boldsymbol{q},\boldsymbol{\sigma}(\boldsymbol{q}))=\sum_{j\in\mathcal{I}}\sigma_j^i(\boldsymbol{q})\lambda_j,\quad i\in\mathcal{I}\ . \tag{5}$$

Our objective is to find the right $\boldsymbol{\sigma}(\boldsymbol{q})\in\mathcal{A}$ for each state $\boldsymbol{q}$ that minimizes (3). Using Little's law, (3) becomes

$$\frac{|\,E\left[\boldsymbol{Q}\right]|}{|\,\boldsymbol{\lambda}\,|}+\sum_{j\in\mathcal{I}}\sum_{i\in\mathcal{I}}\bar{\eta}_j^i+\sum_{j\in\mathcal{I}}\sum_{i\in\mathcal{I}}w_j^i\cdot\bar{\phi}_j^i\ , \tag{6}$$

where $|\cdot|$ denotes the $L_1$ norm.

Under a fixed policy $\boldsymbol{\sigma}(\boldsymbol{q})$, the process $(\boldsymbol{Q}(t),t\geq 0)$ is a Markov process on $\mathbb{Z}_+^N$ with state-dependent (depends on both channel and queue states) transition rate. For convenience, we uniformize $\boldsymbol{Q}(t)$ following [27]. For any $\boldsymbol{q}\in\mathbb{Z}_+^N$, we make the following definitions:

$$D_i\boldsymbol{q}\equiv\max(\boldsymbol{0},\boldsymbol{q}-\boldsymbol{e}_i)\ ,\qquad A_i\boldsymbol{q}\equiv\boldsymbol{q}+\boldsymbol{e}_i\ , \tag{7}$$

where $\boldsymbol{e}_i$ is a $1\times N$ zero-valued vector except the $i^{th}$ element is 1 and max is the element-wise maximum operation. The $D_i$ in (7) denotes a file is successfully transmitted from the BS to user $i$, and $A_i$ means a file arrives to $Q_i$ at the BS.

Let $\varphi\geq|\boldsymbol{\lambda}|+\max_{\boldsymbol{q}}|\boldsymbol{\mu}(\boldsymbol{q})|$. Let $\tau_k$ denote the time of the $k^{th}$ transition of $\boldsymbol{Q}(t)$ and $\tau_0=0$. Also, let $\boldsymbol{Q}_k=\lim_{t\downarrow\tau_k}\boldsymbol{Q}(t)$. Then, under policy $\boldsymbol{\sigma}(\boldsymbol{q})$, the process $\boldsymbol{Q}(t)$ can be viewed as having a state-independent event transition rate of $\varphi$, and the transition probabilities are given by

$$P\big(\boldsymbol{Q}_{k+1}=A_i\boldsymbol{q}\mid\boldsymbol{Q}_k=\boldsymbol{q}\big)=\lambda_i'(\boldsymbol{q},\boldsymbol{\sigma}(\boldsymbol{q}))/\varphi\ , \tag{8}$$

$$P\big(\boldsymbol{Q}_{k+1}=D_i\boldsymbol{q}\mid\boldsymbol{Q}_k=\boldsymbol{q}\big)=\mu_i(\boldsymbol{q})/\varphi\ , \tag{9}$$

$$P\big(\boldsymbol{Q}_{k+1}=\boldsymbol{q}\mid\boldsymbol{Q}_k=\boldsymbol{q}\big)=1-\big(|\boldsymbol{\lambda}|+|\boldsymbol{\mu}(\boldsymbol{q})|\big)/\varphi. \tag{10}$$

Moreover, we define the function $\boldsymbol{f}(\cdot)$ as

$$\boldsymbol{f}\left(\boldsymbol{q},\boldsymbol{\sigma}(\boldsymbol{q})\right)=\sum_{j\in\mathcal{I}}\sum_{i\in\mathcal{I}}\sigma_j^i(\boldsymbol{q})\cdot\left[\eta_j^i(\theta_j)+w_j^i\cdot\phi_j^i(\theta_j)\right]\ . \tag{11}$$

The cost (our objective in (6)) under policy $\boldsymbol{\sigma}$ over $[0,\tau_k)$ when starting from an initial queue state $\boldsymbol{q}$ is

$$\mathbb{E}_{\boldsymbol{q}}^{\boldsymbol{\sigma}}\left[\int_0^{\tau_k}\left[\frac{|\,\boldsymbol{Q}(t)\,|}{|\,\boldsymbol{\lambda}\,|}+\boldsymbol{f}\big(\boldsymbol{Q}(t),\boldsymbol{\sigma}(\boldsymbol{Q}(t))\big)\right]dt\right]\ , \tag{12}$$

which, by ignoring the constant multiplier $\varphi^{-1}$, is equal to:

$$V_k^{\boldsymbol{\sigma}}(\boldsymbol{q})\equiv\mathbb{E}_{\boldsymbol{q}}^{\boldsymbol{\sigma}}\left[\sum_{l=0}^{k-1}\left(\frac{|\,\boldsymbol{Q}_l\,|}{|\,\boldsymbol{\lambda}\,|}+\boldsymbol{f}\left(\boldsymbol{Q}_l,\boldsymbol{\sigma}(\boldsymbol{Q}_l)\right)\right)\right]\ . \tag{13}$$

Then the average cost under policy $\boldsymbol{\sigma}$ when starting from state $\boldsymbol{q}$ is given as follows:

$$J_{\boldsymbol{q}}^{\boldsymbol{\sigma}}=\lim_{k\to\infty}\sup\frac{1}{k}V_k^{\boldsymbol{\sigma}}(\boldsymbol{q})\ . \tag{14}$$

The objective function given in (6) seeks to find the minimal average cost and the corresponding optimal control, which fits the classical dynamic programming (refer to Sec. 7.4 of [27]). Under all possible dispatching probabilities $\boldsymbol{\sigma}(\boldsymbol{q})\in\mathcal{A}$, the minimal average cost $J^*$ is well-defined, independent of the initial state $\boldsymbol{Q}(0)$, and satisfies Bellman's equation:

$$\begin{aligned}J^*=&\min_{\boldsymbol{\sigma}(\boldsymbol{q})\in\mathcal{A}}\left\{\frac{|\,\boldsymbol{q}\,|}{|\,\boldsymbol{\lambda}\,|}+\boldsymbol{f}(\boldsymbol{q},\boldsymbol{\sigma}(\boldsymbol{q}))\right.\\&\left.+\mathbb{E}^{\boldsymbol{\sigma}(\boldsymbol{q})}\big\{\big[h\big(\boldsymbol{Q}_{k+1}\big)-h\big(\boldsymbol{Q}_k\big)\big]\mid\boldsymbol{Q}_k=\boldsymbol{q}\big\}\right\}\\=&\frac{|\,\boldsymbol{q}\,|}{|\,\boldsymbol{\lambda}\,|}+\sum_{i\in\mathcal{I}}\frac{\mu_i(\boldsymbol{q})}{\varphi}\Big[h\big(D_i\boldsymbol{q}\big)-h\big(\boldsymbol{q}\big)\Big]+\min_{\boldsymbol{\sigma}(\boldsymbol{q})\in\mathcal{A}}\sum_{j\in\mathcal{I}}\sum_{i\in\mathcal{I}}\\&\frac{\lambda_j\sigma_j^i(\boldsymbol{q})}{\varphi}\Big\{\eta_j^i(\theta_j)+w_j^i\cdot\phi_j^i(\theta_j)+\big[h\big(A_i\boldsymbol{q}\big)-h\big(\boldsymbol{q}\big)\big]\Big\}\ ,\end{aligned} \tag{15}$$

where $h(\boldsymbol{q})=J(\boldsymbol{q})-J(\boldsymbol{q}_s)$ is a relative cost function with $\boldsymbol{q}_s$ being a reference state. The optimal dispatching policy $\boldsymbol{\sigma}^*(\boldsymbol{q})$ that minimizes (15) can be calculated through methods such as the value iteration or policy iteration from the dynamic programming framework [27].

## V. Properties of the Optimal Policy

In this section, we present some properties of the optimal dispatching policy $\boldsymbol{\sigma}^*(\boldsymbol{q})$ when the local communication among users is not congested. The cellular link has a Rayleigh fading channel and the Signal-to-Noise-Ratio (SNR) is assumed to be constant during a time slot. We denote the channel bandwidth as $B$, and the distance between user $i$ and the BS as $d_i$, both of which affect users' data rate. For the channel being in state $k \in \mathcal{S}$, the data rate $R_i^k$ is given by the Shannon formula with a 3dB SNR loss (to model achievable data rate):

$$R_i^k = B \cdot \log_2(1 + \mathrm{SNR}_i(k)/2) \ . \qquad (16)$$

The settings of channel parameters, mean file size and additional power expenditure for spreading files are presented in Sec. VII. Note that the properties of the optimal policy are not sensitive to these settings.

### *A. Restricting the Optimal Policy Space*

The following theorem guarantees that the optimal value of the objective function can be achieved by non-randomized policies that apply deterministic rules for dispatching the arrivals at a given system state. This reduces the computational effort required to compute an optimal dispatching policy, and allows us to use value iteration in the sequel to study the structure of the optimal policy.

*Theorem 1:* There exists an optimal dispatching policy $\boldsymbol{\sigma}^*(\boldsymbol{q})$ such that each element ${\sigma^*}_j^i \in \{0, 1\}$.

*Proof:* From Sec. IV, we know that a dispatching policy that minimizes (15) is an optimal policy. To minimize (15), it is sufficient to minimize the last "min" part. Since the dispatching rule used by each user is chosen independent of the others, we only have to minimize each part within the first sum, namely

$$\min_{\boldsymbol{\sigma}(\boldsymbol{q}) \in \mathcal{A}} \sum_{i \in \mathcal{I}} \sigma_j^i(\boldsymbol{q}) \Big\{ \eta_j^i(\theta_j) + w_j^i \cdot \phi_j^i(\theta_j) + \big[h\big(A_i \boldsymbol{q}\big) - h\big(\boldsymbol{q}\big)\big] \Big\},$$

where $i, j \in \mathcal{I}$. Below we consider a particular value of $j$. Under queue state $\boldsymbol{q}$, we denote:

$$\alpha_i \equiv \sigma_j^i(\boldsymbol{q}), \quad \beta_i \equiv \eta_j^i(\theta_j) + w_j^i \cdot \phi_j^i(\theta_j) + \big[h\big(A_i \boldsymbol{q}\big) - h\big(\boldsymbol{q}\big)\big]$$

Furthermore, let $\boldsymbol{\alpha} = \{\alpha_i, i \in \mathcal{I}\}$ denote a stochastic vector. To prove the theorem, we have to show that for a given $\boldsymbol{\beta}$, the minimal value of $\boldsymbol{\alpha} \cdot \boldsymbol{\beta}$ can be achieved when $\alpha_{i^*} = 1$, where $i^* \in \arg\min_{i \in \mathcal{I}} \{\beta_i\}$, augmented with a tie-breaking rule. For $i^*$ and $\forall i \in \mathcal{I}$, we have

$$1 \cdot \beta_{i^*} = \alpha_{i^*}\beta_{i^*} + \sum_{i \neq i^*} \alpha_i \beta_{i^*} \leq \alpha_{i^*}\beta_{i^*} + \sum_{i \neq i^*} \alpha_i \beta_i \qquad (17)$$

Therefore when the element $\alpha_{i^*}$ of vector $\boldsymbol{\alpha}$ is set to 1 and all the other elements of $\boldsymbol{\alpha}$ are set to 0, the minimal value of $\boldsymbol{\alpha} \cdot \boldsymbol{\beta}$ is achieved, which proves the theorem. ∎

### *B. A Two-user System*

From Theorem 1, we know that under the two-user model and at any queue state $\boldsymbol{q}$, there are three reasonable controls: *i)* $\sigma(\boldsymbol{q}) = [1\ 0; 0\ 1]$; *ii)* $\sigma(\boldsymbol{q}) = [1\ 0; 1\ 0]$; *iii)* $\sigma(\boldsymbol{q}) = [0\ 1; 0\ 1]$. In the rest of this paper, we refer to these controls as *no re-routing*, $U_2 \to U_1$ and $U_1 \to U_2$, respectively. The optimal dispatching policies evaluated numerically under different scenarios are shown in Fig. 3. The axes correspond to the number of files in the users' queues, and the figure depicts the optimal dispatching strategy at each state. From this figure, we observe the following properties:

**Existence of switching curves:** As Fig. 3 shows, the optimal policy in all the above cases consists of a set of switching curves, i.e., the policy is transition monotone [28]. Here, switching curves refer to the boundaries between contiguous regions where the same control is used in each state of the region. We conjecture that an optimum policy can be described by threshold values $q_2^a$ and $q_2^b$ corresponding to each value of $q_1$, such that

$$\sigma^*(q_1, q_2) = \begin{cases} U_1 \to U_2, & \text{if } q_2 \leq q_2^a \\ \textit{No re-routing}, & \text{if } q_2^a \leq q_2 \leq q_2^b \\ U_2 \to U_1, & \text{if } q_2 \geq q_2^b \end{cases}$$

For the two-user homogeneous scenario under a two-state channel model with delay-optimal scheduling policy, we can prove the optimal policy indeed possesses this structure, as shown in the following theorem:

*Theorem 2:* There exists an optimal dispatching policy that has switching curves (transition monotone), under the two-user homogeneous scenarios under a two-state channel model (*on/off*) with delay-optimal scheduling policy.

We introduce some notations and the value iteration method. For the two-state channel (*on/off*), we denoted the probability that user $i$'s channel is *on* as $p_i$, $i \in \{1, 2\}$. Since we consider the homogeneous scenario, we have

$$\eta \cdot z_j^i = \eta_j^i(\theta_j), \quad \phi \cdot z_j^i = \phi_j^i(\theta_j), \quad \forall i, j \in \{1, 2\},$$

where $z_j^i = 1$ if $j \neq i$ and $z_j^i = 0$ if $j = i$.

The value iteration is widely used to solve the Bellman equation. The most used version of the value iteration method for the average cost problem is to select an initial state and generate successively the corresponding optimal $k$-stage cost $J_k(\boldsymbol{q})$. As shown in [27], the ratios $J_k(\boldsymbol{q})/k$ converges to the optimal average cost per stage $J^*$ as $k \to \infty$. Therefore, we can use induction upon the value iteration method to prove our theorem, as the author does in [27]. We first define the relative value $h_k(\boldsymbol{q})$ at stage $k$ given by

$$h_k(\boldsymbol{q}) = J_k(\boldsymbol{q}) - J_k(\boldsymbol{q}_s) \qquad (18)$$

where $\boldsymbol{q}_s$ is a reference state. Moreover, we define $\Delta_k(\boldsymbol{q})$ as follows:

$$\begin{aligned} \Delta_k(\boldsymbol{q}) =& h_k(\boldsymbol{q} + \boldsymbol{e}_1) - h_k(\boldsymbol{q} + \boldsymbol{e}_2) \\ =& J_k(\boldsymbol{q} + \boldsymbol{e}_1) - J_k(\boldsymbol{q}_s) - (J_k(\boldsymbol{q} + \boldsymbol{e}_2) - J_k(\boldsymbol{q}_s)) \\ =& J_k(\boldsymbol{q} + \boldsymbol{e}_1) - J_k(\boldsymbol{q} + \boldsymbol{e}_2) \end{aligned} \qquad (19)$$

The following lemma supports the proof of Theorem 2.

*Lemma 1:* The $\Delta_k(\boldsymbol{q})$ is monotonically non-decreasing in $q_1$ for each fixed $q_2$, where $\boldsymbol{q} \equiv \{q_1, q_2\}, q_1, q_2 \in \mathbb{Z}_+$.

The proof of Lemma 1 is presented in the Appendix.

*Proof of Theorem 2:* We consider the optimal dispatching policy characterized by Theorem 1. To show the optimal policy has switching curves for the two-user homogeneous scenarios, it is sufficient to show that $\Delta_k(\boldsymbol{q})$ is monotonically non-decreasing in $q_i$ for each fixed $q_j$, $i, j \in \{1, 2\}$ and $i \neq j$ [27], which can be obtained from Lemma 1. ∎

**Performance vs. Energy consumption:** Choosing a weight of **0** implies that the optimal policy is one which minimizes

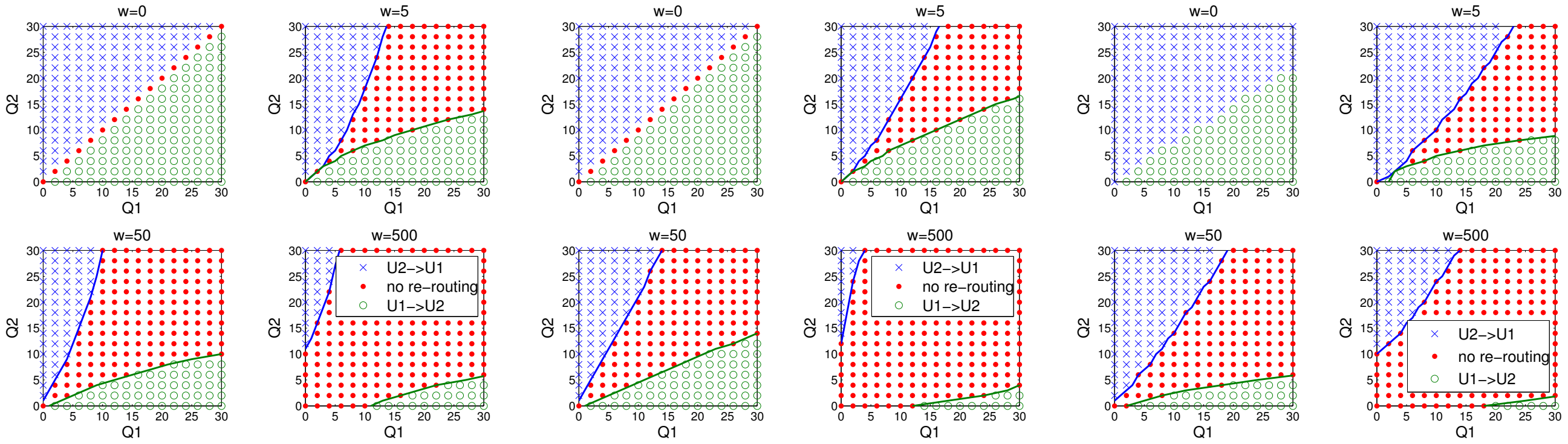

(a) $\boldsymbol{\lambda} = \mathbf{0.2}$ arrivals/sec; (b) $\boldsymbol{\lambda} = \mathbf{0.2}$ arrivals/sec; (c) $\boldsymbol{\lambda} = \{0.19, 0.21\}$ arrivals/sec;

Fig. 3: Optimal dispatching strategy as a function of queue backlogs: (a) Homogeneous scenario ($\boldsymbol{d} = \mathbf{100}$ m), greedy scheduler; (b) Homogeneous scenario ($\boldsymbol{d} = \mathbf{100}$ m), the log rule scheduler; (c) Heterogeneous scenario ($\boldsymbol{d} = \{92, 100\}$ m), greedy scheduler.

average file transfer delay. In the case of the homogeneous scenarios of Fig. 3 (a) and (b), this corresponds to dispatching arrivals to the shortest queue, as described in [21]. In the heterogeneous case of Fig. 3 (c), arrivals are dispatched rather to the queue with less backlog, taking into account the difference in average service rates. A higher value of the weight, $\boldsymbol{w}$, implies that delay performance is sacrificed in order to reduce the excess power expenditure due to traffic spreading. We observe that the regions corresponding to *re-routing areas* ($U_1 \rightarrow U_2$ and $U_2 \rightarrow U_1$) diminish progressively as the weight attached to power expenditure increases. At very high values of $\boldsymbol{w}$, traffic spreading is initiated only when the imbalance between the user queues is very large.

**Switching curve shape:** We observe from the results in Fig. 3 that the level of imbalance between the queues that is required for arrival re-routing to be the optimal strategy increases as the overall backlog increases. For instance, the threshold on the queue length of $U_2$ beyond which arrivals are re-routed to $U_1$ appears to be a convex, increasing function of the backlog in $Q_1$. The intuition behind this is that when the backlog in both queues is large, the time interval for a queue to empty out is likely to be long, and the shorter queue might yet see many arrivals even without re-routing. In such a case, the gain from dispatching requests to other users to balance the queues does not justify the associated power expenditure.

**Dispatching as a function of the scheduling policy:** We observe that the optimal dispatching policy under the log rule scheduler (Fig. 3(b)) consists of switching curves that favor more re-routing, especially at larger queue-lengths. For example, the threshold $q_2^b$ prompting re-routing is lower, and does not increase as rapidly with total queue length as under the greedy scheduler (Fig. 3(a)). The log rule scheduler itself reacts to imbalance in queues, sacrificing opportunistic gain in order to balance the queues. The optimal dispatcher takes this into account, resulting in the above policy. However, as we will see in the sequel, the dispatcher complements the log rule scheduler well and the combination does achieve better performance at lower power expenditure compared to the queue-unaware greedy scheduler.

**Impact of the request arrival rate:** Switching curves of the optimal dispatching policy under the greedy scheduler for different request arrival rates are shown in Fig. 4. We can observe that with the increase of arrival rate, the re-routing areas decrease. One reason for this is that at lower arrival rates, the shorter queue could be emptied before the next arrival to either user. Thus, the optimal dispatcher is more aggressive at re-routing requests even when queue lengths are larger.

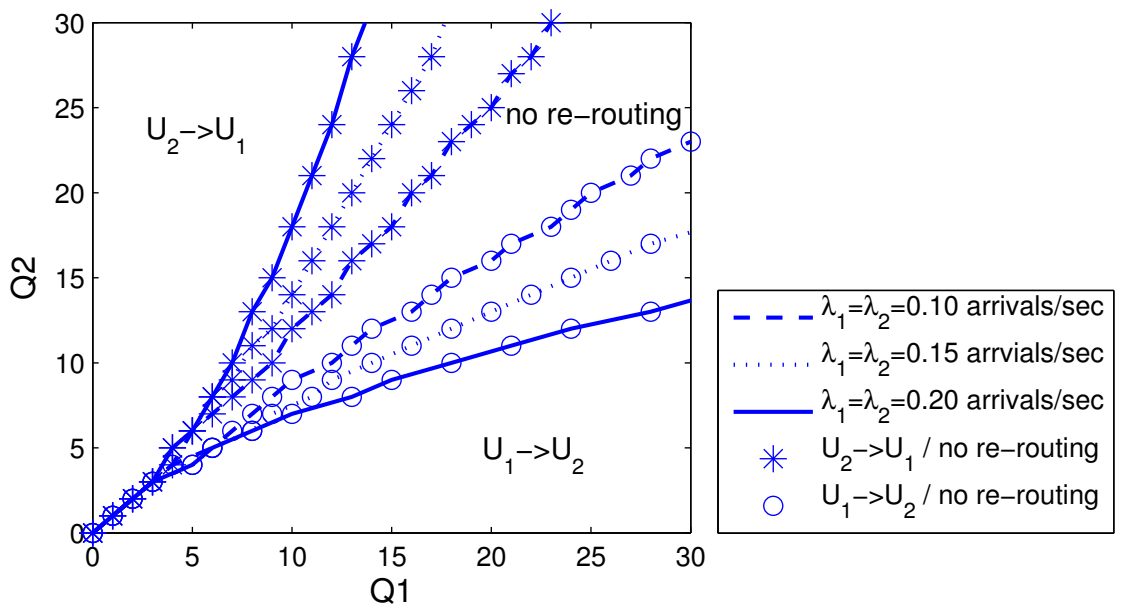

Fig. 4: Switching curves of the optimal dispatching policy under different request arrival rates where $\boldsymbol{d} = \mathbf{100}$ m and $\boldsymbol{w} = \mathbf{5}$.

## VI. A Heuristic Algorithm for Multi-user System

In scenarios with many users, numerically computing the optimal dispatching policy by solving the dynamic programming formulation (15) becomes intractable. Thus we propose a heuristic algorithm to determine the dispatching policy for a multi-user system that uses the dynamic programming solution of two-user scenarios as a building block. Here, we focus on currently deployed queue-unaware schedulers such as the greedy policy. The heuristic used to compute dispatching decisions at user $i$ is specified below in Algorithm 1. We denote by $dp(\tilde{\boldsymbol{\lambda}}, \tilde{\boldsymbol{\mu}}, \tilde{\boldsymbol{\phi}})$, the optimal dispatching policy for the two-user model, where $\tilde{\boldsymbol{\lambda}} = \{\tilde{\lambda}_1, \tilde{\lambda}_2\}$ is the vector of arrival rates , $\tilde{\boldsymbol{\mu}}(\tilde{\boldsymbol{q}}) = \{\tilde{\boldsymbol{\mu}}_1(\tilde{\boldsymbol{q}}), \tilde{\boldsymbol{\mu}}_2(\tilde{\boldsymbol{q}})\}, \tilde{\boldsymbol{q}} \in \mathbb{Z}_+^2$ specifies the queue-state dependent service rates and $\tilde{\boldsymbol{\phi}} = \{\tilde{\phi}_1^2(\theta_1), \tilde{\phi}_2^1(\theta_2)\}$ is the re-routing power expenditure.

The proposed heuristic considers two options for a new request, i.e., forwarding it directly to the BS or dispatching it to the user with the least amount of work (workload) in its

**Algorithm 1** A Heuristic Algorithm for N-user System

// Heuristic to dispatch a new arrival at user $i$.

// **Definitions:** 1) $\boldsymbol{e}_l \equiv$ a $1 \times N$ zero-valued vector except the $l^{th}$ element is 1; 2) $\mathbf{1}_N \equiv \sum_{l=1}^{N} \boldsymbol{e}_l$.

**Input:** $\boldsymbol{\lambda}$ **:** a $1 \times N$ vector of arrival rates

$\boldsymbol{q}$ : a $1 \times N$ vector of queue state

$\boldsymbol{\mu}(\boldsymbol{q})$: a matrix of queue-state dependent service rates

**Output:** dispatching decision at user $i$

1: $j \leftarrow \arg\min_{l \in \mathcal{I}} \{q_l/\mu_l(\boldsymbol{e}_l)\}$.
2: **if** $j \neq i$ **then**
3: $\quad \mathcal{Y}_S \leftarrow \{j\}$; $\mathcal{Y}_B \leftarrow \mathcal{I} \setminus \mathcal{Y}_S$; $\mathcal{Y}_P \leftarrow \emptyset$
4: $\quad \tilde{q}_1 \leftarrow \sum_{l \in \mathcal{I} \setminus \mathcal{Y}_S} q_l$; $\tilde{q}_2 \leftarrow q_j$
5: $\quad \tilde{\boldsymbol{\mu}}_1(\tilde{\boldsymbol{q}}) \leftarrow \begin{cases} 0, & \tilde{q}_1 = 0 \\ \sum_{l \neq j} \mu_l(\mathbf{1}_N - \boldsymbol{e}_j), & \tilde{q}_1 > 0, \tilde{q}_2 = 0 \\ \sum_{l \neq j} \mu_l(\mathbf{1}_N), & \tilde{q}_1 > 0, \tilde{q}_2 > 0 \end{cases}$
6: $\quad \tilde{\boldsymbol{\mu}}_2(\tilde{\boldsymbol{q}}) \leftarrow \begin{cases} 0, & \tilde{q}_2 = 0 \\ \mu_j(e_j), & \tilde{q}_1 = 0, \tilde{q}_2 > 0 \\ \mu_j(\mathbf{1}_N), & \tilde{q}_1 > 0, \tilde{q}_2 > 0 \end{cases}$
7: $\quad$ **while** $\mathcal{Y}_B \neq \emptyset$ **do**
8: $\quad\quad i^* \leftarrow \arg\max_{l \in \mathcal{Y}_B} \{q_l/\mu_l(\boldsymbol{e}_l)\}$
9: $\quad\quad \tilde{\lambda}_1 \leftarrow \sum_{l \in \mathcal{I} \setminus \mathcal{Y}_S} \lambda_l$; $\tilde{\lambda}_2 \leftarrow \sum_{l \in \mathcal{Y}_S} \lambda_l$
10: $\quad\quad \tilde{\boldsymbol{\phi}} \leftarrow \{\phi^j_{i^*}(\theta_{i^*}), \phi^{i^*}_j(\theta_j)\}$
11: $\quad\quad \boldsymbol{\sigma} \leftarrow dp(\tilde{\boldsymbol{\lambda}}, \tilde{\boldsymbol{\mu}}, \tilde{\boldsymbol{\phi}})$
12: $\quad\quad$ **if** $\boldsymbol{\sigma}(\tilde{q}_1, \tilde{q}_2) = U_1 \rightarrow U_2$ **then**
13: $\quad\quad\quad$ **if** $i^* = i$ **then**
14: $\quad\quad\quad\quad$ **return** dispatch the new request to user $j$
15: $\quad\quad\quad$ **else**
16: $\quad\quad\quad\quad \mathcal{Y}_B \leftarrow \mathcal{Y}_B \setminus i^*$; $\mathcal{Y}_S \leftarrow \mathcal{Y}_S \cup i^*$
17: $\quad\quad\quad$ **end if**
18: $\quad\quad$ **else if** $i^* = i$ **then**
19: $\quad\quad\quad$ **return** send the new request directly to the BS
20: $\quad\quad$ **else**
21: $\quad\quad\quad \mathcal{Y}_B \leftarrow \mathcal{Y}_B \setminus i^*$; $\mathcal{Y}_P \leftarrow \mathcal{Y}_P \cup i^*$
22: $\quad\quad$ **end if**
23: $\quad$ **end while**
24: **else**
25: $\quad$ **return** send the new request directly to the BS
26: **end if**

BS queue. To this end, all users other than the one with the least workload are treated as a single combined user, and their queues are also treated as a single combined queue. A series of two-user dynamic programming formulations are solved, where the two users are the combined user and the user with least workload. In order to determine the parameters of the two-user dynamic program, we map states where the combined queue is non-empty to states where all the component queues are non-empty in the multi-user system. The service rate of the combined queue at a state is calculated as the sum of the service rates of the component queues in the corresponding state (steps 4-6).

The users are examined in sequence, and the dispatching strategy is decided in order of decreasing workload. The arrival rates to the combined queue reflects the dispatching decisions made at all the users with higher workload than the one currently under consideration (step 9). The dynamic programming solution is computed taking into account the power expenditure associated with dispatching from the current user to the one with least workload. The state considered in the reduced dynamic program is always one where the combined user queue length is the sum of the queue lengths of the component queues. Arrivals to the current user are dispatched to the user with least workload in the multi-user system if the optimal policy in the reduced scenario is to re-route from the combined queue to the other. Note that the worst-case time-complexity of the above heuristic to obtain the dispatching decision for a new request is $O(N-1)$.

## VII. Performance Evaluation

In this section, we evaluate our proposed traffic spreading policy through simulations and demonstrate the tradeoff between performance improvement and additional power expenditure resulting from the dynamic programming formulation and our multi-user heuristic. We assume the mean file size $\theta_j$=1MB, under which the additional power expenditure $\phi^i_j(\theta_j)$ is 1 Joule for all $i \neq j, i, j \in \mathcal{I}$. The settings of the channel parameters are listed in Table I. We ignore the user-user delay (i.e., $\eta^i_j = 0, \forall i, j \in \mathcal{I}$) (thus the local communication among users is uncongested) since the data rate of user-user link would be much higher than the cellular link in reality [29]. We estimate the average file transfer delay and additional power expenditure within a relative error of 2%, at a confidence interval of 95%.

TABLE I: Channel parameters

| Parameters | Value |
|---|---|
| Bandwidth (Min bandwidth in LTE) | 1.4 MHz |
| BS Tx power spectral density | 0.1/1.4 W/MHz |
| Noise spectral density | $10^{-8}$/1.4 W/MHz |
| Path loss exponent (Urban Area) | 3 |
| Slot time | 10 ms |
| Doppler shift (ITU Pedestrian A) | 5 Hz |

To evaluate the proposed traffic spreading, we compare it with two dispatching policies:

*1) No re-routing:* Under this dispatching policy, when a request of user $i$ is generated, user $i$ sends it directly to the BS. Thus there is no additional power expenditure.

*2) Join the Shortest Queue (JSQ):* Under this dispatching policy, when there is a new request generated by user $i$, user $i$ sends it to user $j$ that has the least amount of work left. If $i \neq j$, then additional power expenditure occurs.

We also propose a lower bound on the average file transfer delay. This bound is obtained from an algorithm consisting of: $i$) using our proposed traffic spreading algorithm at the users; $ii$) modifying the original BS scheduling policy according to Algorithm 2. The motivation behind the lower bound is when the queue of the user with the highest instantaneous data rate is empty, the opportunistic gain will be lost. To recoup the opportunistic gain, the BS can re-route data from other queues to the user with the highest instantaneous data rate. Besides, re-routing a user's own data from other queues saves re-routing cost, while re-routing a user's data to another user requires re-routing cost. Algorithm 2 takes these into account and the re-routing at the BS is forced not to bring obvious additional re-routing cost.

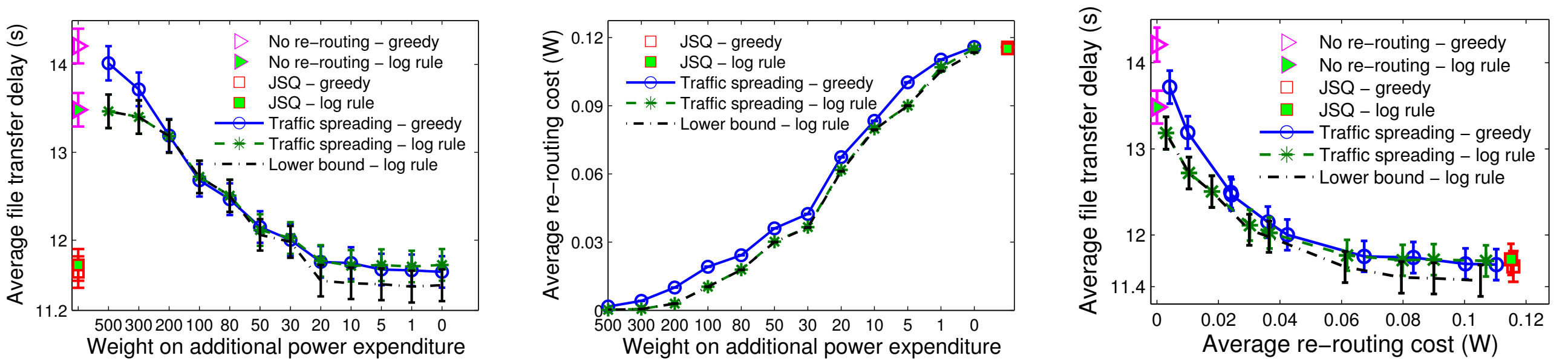

(a) Average delay versus weight (b) Average re-routing cost versus weight (c) Average delay versus re-routing cost

Fig. 5: Performance under two-user homogeneous scenarios where $\boldsymbol{\lambda} = \mathbf{0.2}$ arrivals/sec and $\boldsymbol{d} = \mathbf{100}$ m.

**Algorithm 2** A lower bound of our proposed algorithm

// A modification to the original BS scheduling policy.
// **Definitions:** $\theta_{bs}$: record the re-routed data at the BS.
**Input:** $\boldsymbol{R}(t) = \{R_i^{c_i(t)}, i \in \mathcal{I}, c_i(t) \in \mathcal{S}\}$ : a $1 \times N$ vector of channel-state dependent service rate at time slot $t$
$\boldsymbol{q}(t)$ : a $1 \times N$ vector of queue state
$\boldsymbol{b}(t)$ : a $1 \times N$ vector of the original owner of the head-of-line file of each queue at time slot $t$

$j \leftarrow \arg\max_{i \in \mathcal{I}} R_i^{c_i(t)}$.
**if** $q_j(t)$ is empty **then**
  **if** $\mathcal{K} \equiv \{i : b_i(t) = j, i \in \mathcal{I}\}$ and $\mathcal{K}! = \emptyset$ **then**
    schedule a user $k \in \mathcal{K}$
    $\theta_{bs} = \theta_{bs} -$ served data from user $k$'s queue
    **return**
  **else if** $\mathcal{K} \equiv \{i : b_i(t)! = i, i \in \mathcal{I}\}$ and $\mathcal{K}! = \emptyset$ **then**
    schedule a user $k \in \mathcal{K}$
    **return**
  **else if** $\theta_{bs} < 0$ **then**
    $k \leftarrow \arg\max_{i \in \mathcal{I}} q_i(t)$
    schedule the user $k$
    $\theta_{bs} = \theta_{bs} +$ served data from user $k$'s queue
    **return**
  **end if**
**end if**
**return** execute the original scheduling algorithm

### A. The Two-user Scenario

**Homogeneous scenarios:** The simulation results under the greedy and log rule scheduling policies are shown in Fig. 5 (a)-(c) for a homogeneous scenario where both users are at exactly the same distance from the BS and have identical traffic demands. Fig. 5 (a) shows that JSQ results in the lowest average delay independent of the scheduling policy, as expected. Under performance-centric case ($\boldsymbol{w} = \mathbf{0}$), traffic spreading can reduce the average delay as much as JSQ does, independent of the scheduling policies. Both JSQ and the traffic spreading have delay improvement up to 18% (greedy) and 14% (log rule) compared to no re-routing. Besides, the gap between our proposed algorithm and the lower bound increases with the decrease of weight. This is because the smaller the weight, the more re-routing at the users will occur. Thus the BS has more opportunities to re-route a user's own data from others' queues to save re-routing cost, and then is able to re-route a user's data to other users. The results in Fig. 5(b) show that these strategies also correspond to the highest power expenditure. Traffic spreading re-routes as much as JSQ does when $\boldsymbol{w} = \mathbf{0}$. Under energy-sensitive cases ($\boldsymbol{w} > \mathbf{0}$), increasing the weight of re-routing cost results in the energy consumption decreasing rapidly along with increasing average file transfer delay.

Moreover, we observe from Fig. 5 (a) that under large weight ($\boldsymbol{w} \geq \mathbf{200}$), the delay performance under the log rule scheduler is better than under the greedy scheduler. However, in the performance-centric scenarios, traffic spreading under the greedy scheduler does achieve similar delay performance. In general, the rate of re-routing and the power expenditure is lower under the log rule scheduler. This is because the log rule scheduling policy already tries to balance the queues, while the greedy scheduler does not.

The tradeoff between the average delay and re-routing cost under traffic spreading can be seen from Fig. 5 (c). We see again that traffic spreading is generally able to achieve the same delay performance at lower power expenditure compared to the greedy scheduler. A very interesting observation indicated by this figure is that most of the delay performance gain can be achieved with a small increase in power expenditure. For example, when $\boldsymbol{w} = \mathbf{0}$, the (maximal) delay performance gain is 18%, and the (maximal) average re-routing cost is $0.12W$. However, when $\boldsymbol{w} = \mathbf{100}$, the delay performance gain is 14% and the re-routing cost is about $0.025W$. This means under the traffic spreading 78% of the maximal performance gain can be achieved with only 20% of the maximal re-routing cost.

The impact of arrival rates on performance is shown in Fig. 6, where we scale $\boldsymbol{\lambda}$, while keeping $\boldsymbol{d}$ and $\boldsymbol{w}$ unchanged. Even at low loads, traffic spreading results in performance gains, with the gains increasing as $\boldsymbol{\lambda}$ increases. For example, when $\boldsymbol{\lambda} = \mathbf{0.175}$ arrivals/sec, the gain is 16%; and it increases to 23% when $\boldsymbol{\lambda} = \mathbf{0.215}$ arrivals/sec. Note that, with the chosen weight, traffic spreading achieves similar delay performance to JSQ at all traffic loads while consuming less than half the extra energy.

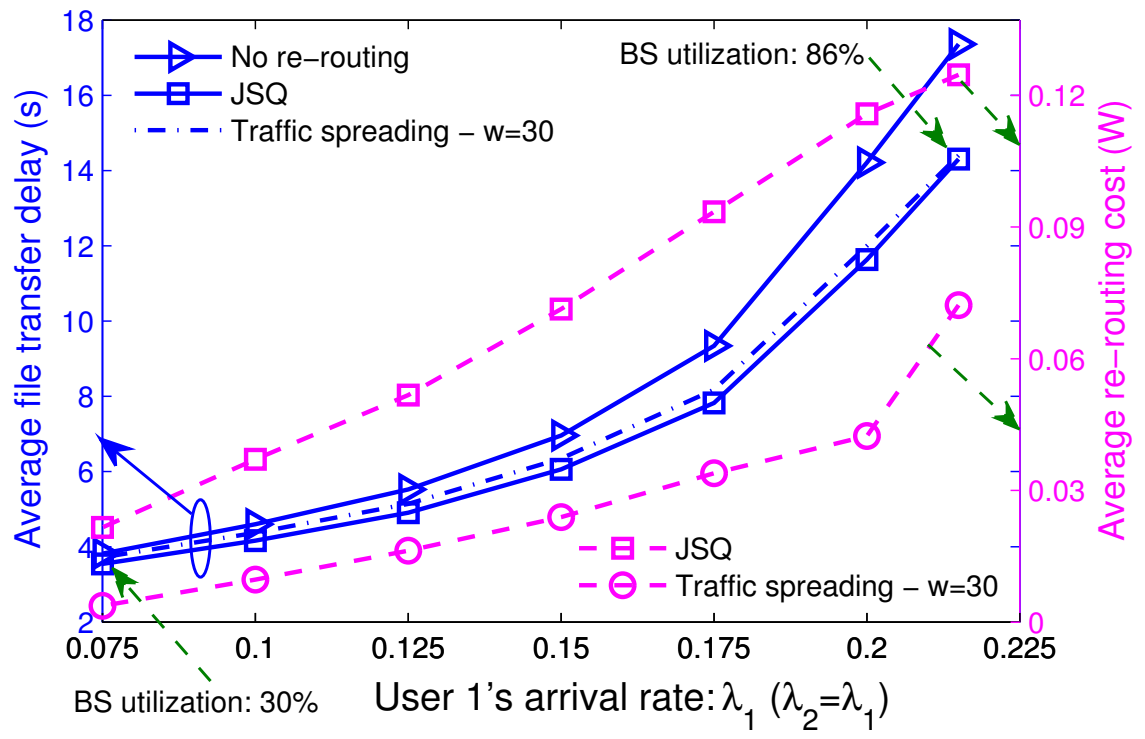

Fig. 6: Performance vs. arrival rate under two-user homogeneous scenarios where $\boldsymbol{d} = \mathbf{100}$ m and $\boldsymbol{w} = \mathbf{30}$ (confidence intervals are not shown for clarity).

**Heterogeneous scenarios:** Fig. 7 depicts the delay performance vs. power expenditure tradeoff achieved by traffic spreading in a scenario where one of the users has lower offered traffic as well as a better average channel to the BS. The maximal delay performance gain under traffic spreading is up to 27% (greedy) and 18% (log rule), compared to no re-routing. Similarly to the homogeneous scenario, the average re-routing power expenditure reduces rapidly as the weight is increased from 0 while the average file transfer delay increases much slower. For example, up to 95% ($\boldsymbol{w} = \mathbf{10}$) of the maximal performance gain can be achieved at only 40% of the maximal re-routing cost. We see again that queue-aware scheduling is indeed beneficial and the combination of the log rule scheduler and traffic spreading is more effective, especially in energy-sensitive scenarios.

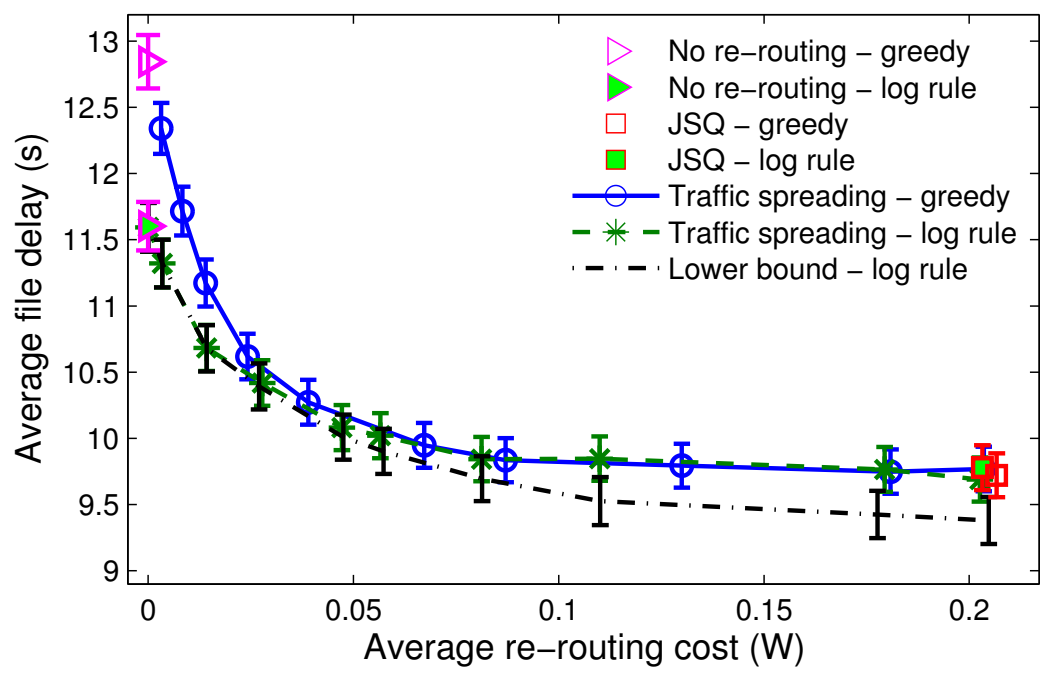

Fig. 7: Performance vs. re-routing cost under two-user heterogeneous scenarios where $\boldsymbol{\lambda} = \{0.19, 0.21\}$ arrivals/sec and $\boldsymbol{d} = \{92, 100\}$ m.

Fig. 8 depicts the overall re-routing rate as well as the split between users. Clearly, the performance gain does not originate from simple relaying. In fact, the user with the better average channel and lower traffic ($U_1$) also forwards traffic to the user with the worse channel ($U_2$). $U_1$ does contribute to the bulk of the performance improvement, however we see that $U_2$ forwards a significant amount of traffic for $U_1$ as well. For example, 40% of the total re-routed traffic is from $U_1$ to $U_2$ in the performance-centric scenario with $\boldsymbol{w} = \mathbf{0}$.

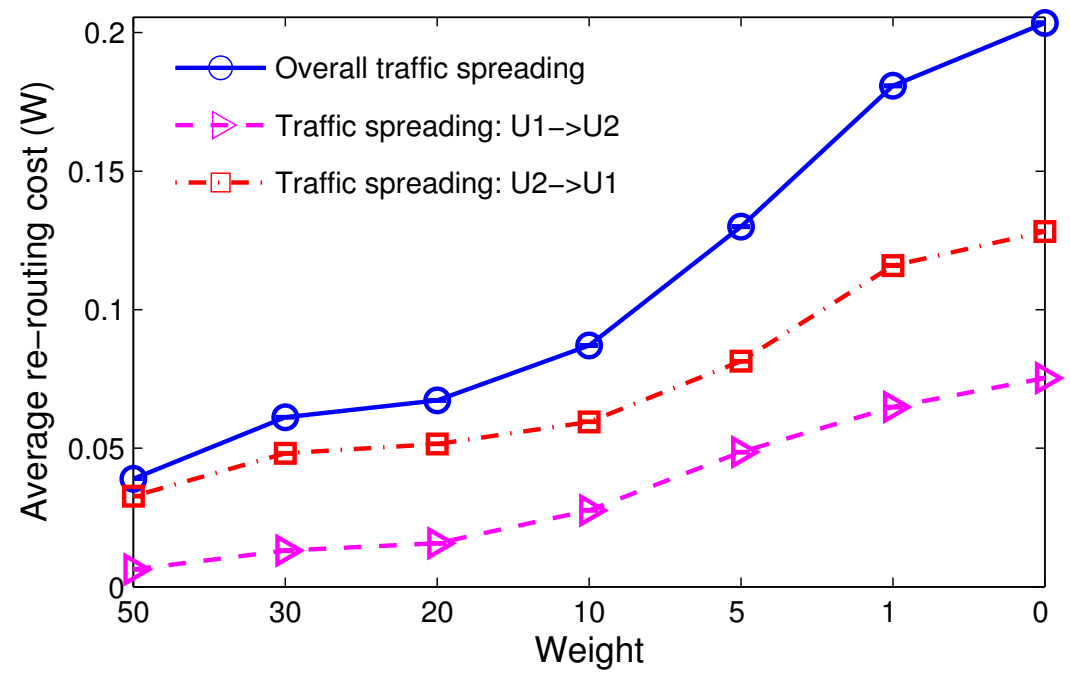

Fig. 8: Average re-routing rate of each user under greedy scheduler where $\boldsymbol{\lambda} = \{0.19, 0.21\}$ arrivals/sec and $\boldsymbol{d} = \{92, 100\}$m.

The traffic spreading algorithm can also adjust the difference on additional power expenditure among users in heterogeneous scenarios, by assigning different weights to different users. Fig. 9 depicts the performance versus average power expenditure as well as the average re-routing rates of users in a heterogeneous scenario, where the weight $w_2^1$ changes between 0 and 500 and the weight $w_1^2$ is fixed to 0 . We can see from Fig. 9 (b) that as $w_2^1$ increases, the difference of re-routing rates between $U_1 \to U_2$ and $U_2 \to U_1$ decreases. The re-routing rate of $U_1 \to U_2$ also decreases with the increase of $w_2^1$. This is because our traffic spreading aims at balancing users' BS queues. As $w_2^1$ increases, the imbalance of queues is enlarged (the average amount of work in $Q_2$ is larger than that in $Q_1$). Therefore, the chance is reduced that $U_2$ helps $U_1$ re-route its requests to the BS, even $w_1^2 = 0$. The average file transfer delay also increases with the increase of $w_2^1$ as expected, as shown in Fig. 9 (a).

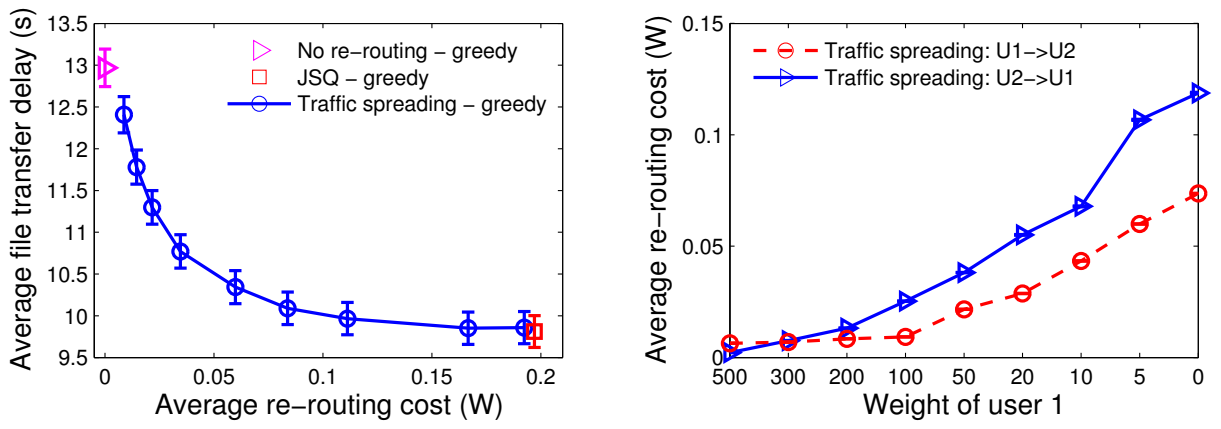

(a) Average file transfer delay versus re-routing cost

(b) Average re-routing cost of users

Fig. 9: Fairness under two-user heterogeneous scenarios where $\boldsymbol{\lambda} = \{0.19, 0.21\}$ arrivals/sec and $\boldsymbol{d} = \{92, 100\}$m.

### B. Multi-user Scenarios

In this subsection, we present the performance evaluation results of the proposed heuristic under the greedy scheduler.

**Dynamic programming vs. heuristic performance:** We first consider a three-user homogeneous scenario. We evaluate the performance of the proposed heuristic against that of the optimal dispatching policy obtained from solving the dynamic programming formulation. We observe from Fig. 10 that the performance of our proposed heuristic is almost as good as that of the optimal dispatching policy, with both able to achieve near identical performance vs. energy consumption tradeoffs. The maximal performance difference between the heuristic and the optimal dispatching policy is less than 2%. The average file transfer delay can be reduced by 18% in this scenario. Similar to the cases considered earlier, we see that up to 60% ($\boldsymbol{w} = \mathbf{30}$) of the maximal performance gain can be achieved at only 20% of the maximal re-routing cost. These results demonstrate that the proposed heuristic is indeed successful in multi-user cases.

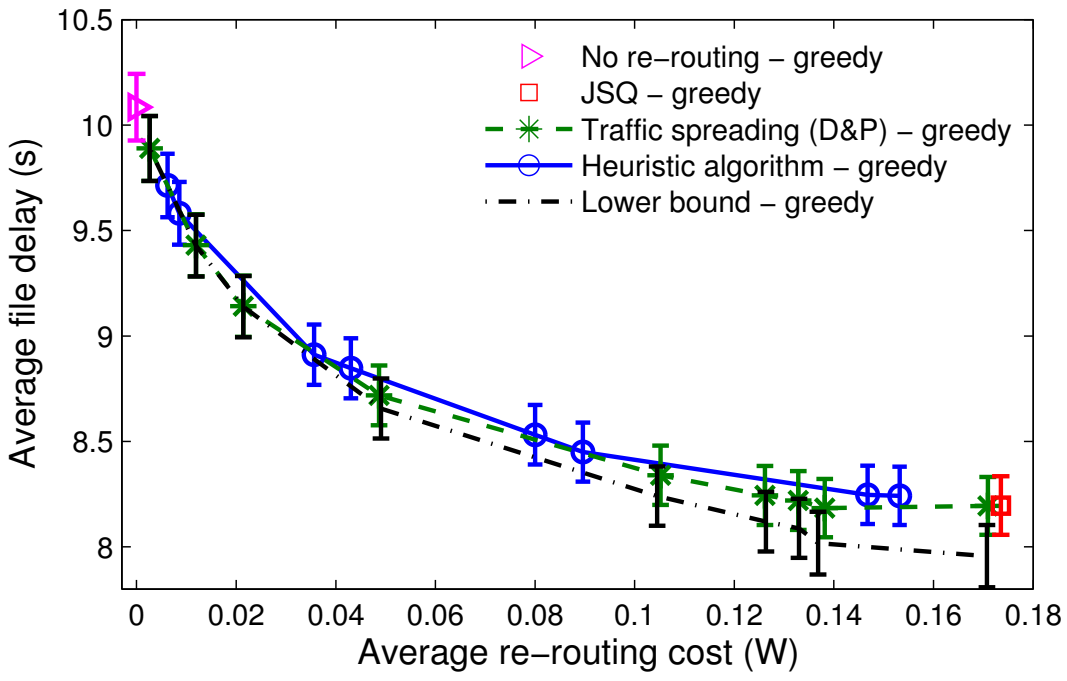

Fig. 10: Dynamic programming vs. heuristic performance under a three-user scenario where $\sum_{i=1}^{3} \lambda_i = 0.4$ arrivals/sec, $\lambda_i = \lambda_j$ and $d_i = 100$m, $i, j \in \{1, 2, 3\}$.

**Performance scaling with number of users:** The scaling of the average file transfer delay with the number of users is shown in Fig. 11 under traffic spreading with different choices of weight, $\boldsymbol{w}$. Here, all users are at the same distance from the BS and offer the same traffic, and the sum arrival rate across all users is fixed to 0.4 arrivals/sec. As we would expect, the average delay decreases with the increase number of users due to the increase in overall opportunistic gain. We can also observe that traffic spreading is able to improve the delay performance by 17% to 19%, compared to no re-routing. As the number of users increases, we observe that the gain from traffic spreading first increases slightly and then decreases as the user population increases further. This is due to the fact that opportunistic gain grows slower than linearly with the size of the user population. Note that even in multi-user scenarios, traffic spreading does result in significant performance gains. The power expenditure trends with increasing weight are similar to those of earlier scenarios, with power expenditure reducing steeply with small increase in file transfer delays when the weight, $\boldsymbol{w}$, is increased (energy consumption curves not shown due to space limitations).

**Users distributed randomly in a cell of radius 100m:** We present simulation results for a scenario where a BS serves users in a service area of radius 100m. We consider instances with four users, each of them distributed uniformly at distances ranging from 10 to 100m (corresponding to average capacity to the BS between 16.0407 to 3.0163Mbps). The rate at which users generate requests is also heterogeneous and chosen uniformly in a range of $0.2 \pm 10\%$ arrivals/sec. We evaluate 50 random instances for each choice of weight (i.e. performance-energy tradeoff), and depict the average as well as the $95^{th}$ and $5^{th}$ percentile of the file transfer delay in Fig. 12. We observe that the average delay performance gain is up to 50%, which is nearly as good as JSQ. This delay performance can be achieved at a re-routing cost that is around half of that under JSQ. Depending on the user requirements, a different tradeoff between power expenditure and delay performance can be chosen. When we focus on the $95^{th}$ percentile, the performance gain observed is up to 56%. This demonstrates that traffic spreading is very helpful in improving user performance in instances where overall system performance is poor, which is a very important practical consideration.

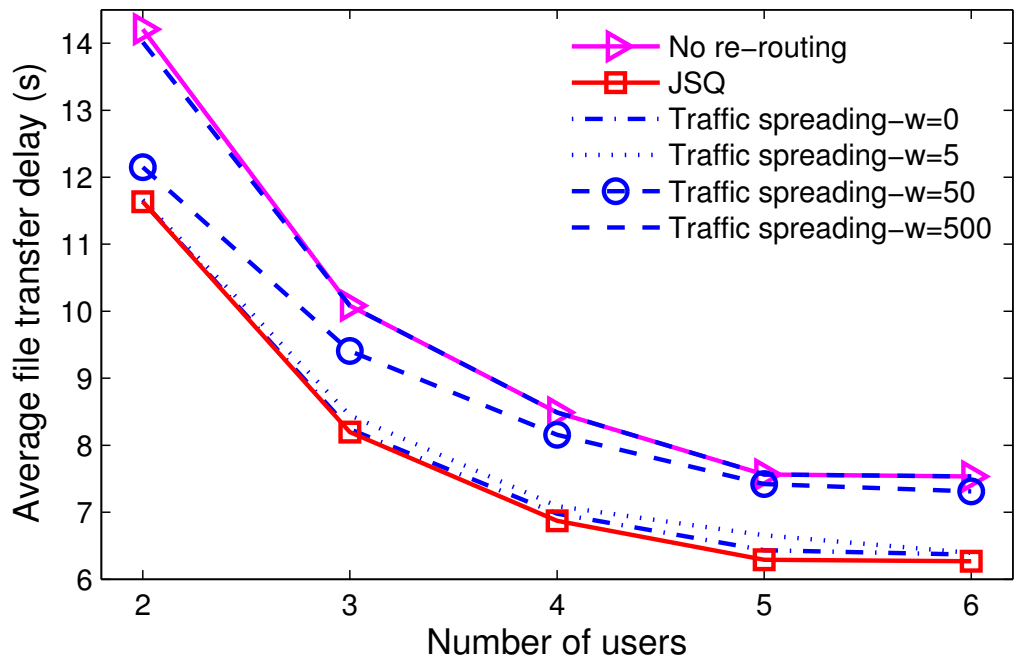

Fig. 11: Performance versus $N$ users where $\sum_{i=1}^{N} \lambda_i = 0.4$ arrivals/sec, $\lambda_i = \lambda_j$ and $d_i = 100$m, $i, j \in \{1, 2, ..., N\}$.

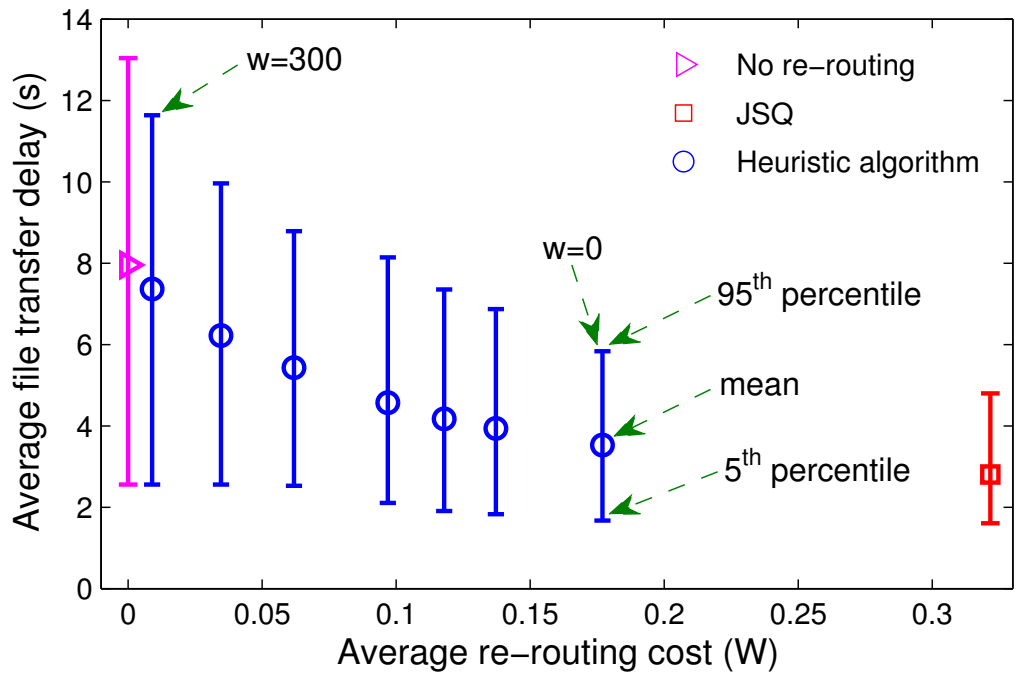

Fig. 12: Performance under four-user heterogeneous scenarios where $\lambda_i = 0.2 \pm 10\%, d_i \in [10, 100], i \in \{1, 2, 3, 4\}$.

## VIII. Traffic Spreading in Large Cells

In this section, we extend our proposed traffic spreading algorithm to large cells, where not all users can communicate with each other. We begin by presenting modifications to the system model and formulation in Sec. III and IV, and provide the simulation results.

### A. Modification to the system model and formulation

To extend the proposed traffic spreading algorithm to large cells, we divide users into clusters where users in the same cluster can communicate with each other directly. As described in Sec. IV, the queue-state-dependent average service rates are indispensable in the dynamic programming formulation. Due

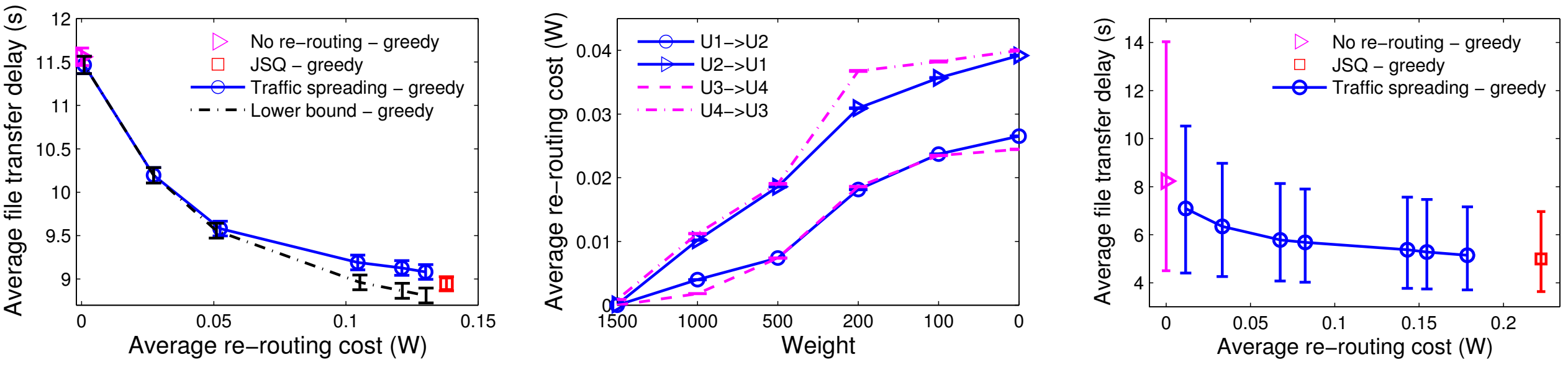

(a) Four-user heterogeneous scenario (b) Average re-routing cost of users (c) Five users randomly distributed in a cell

Fig. 13: Performance in a large cell: $(a)$ Performance under a four-user heterogeneous scenario where $\boldsymbol{\lambda} = \{0.13, 0.12, 0.15, 0.14\}$ arrivals/sec and $\boldsymbol{d} = \{100, 92, 86, 78\}$ m; $(b)$ Average re-routing rate of users under the four-user heterogeneous scenario; $(c)$ Five users randomly distributed in a cell where BS transmission power is 0.3 W/MHz, $\lambda_i = 0.13 \pm 10\%$ arrivals/sec, $d_i \in [50, 200]$ m, $i \in \{1, 2, \cdots, 5\}$.

to the fact that not all users are within the transmission range of each other in a large cell, these service rates can not be calculated accurately at each user, even through users know the scheduling policy employed by the BS. Therefore, we propose a method to estimate the average service rates based on users' statistics in each cluster.

Suppose users in a large cell are divided into $L$ clusters. Let $\mathcal{L}$ denote the set of clusters and $\mathcal{I}_l$ be the set of users in the $l$th cluster. Let $q_{l,i}(t)$ be the queue length of user $i$ and let $s_{l,i}(t)$ be a indicator representing whether user $i$ is served by the BS at slot $t$ or not, where $i \in \mathcal{I}_l, l \in \mathcal{L}$. Moreover, the average time ratio that the BS spends on serving users of cluster $l$ during the time interval $[t_s, t_e]$ is denoted as $\alpha_l$, which is estimated through the following equation

$$\alpha_l = \frac{\sum_{t=t_s}^{t_e} \sum_{i \in \mathcal{I}_l} q_{l,i}(t) \wedge s_{l,i}(t)}{t_e - t_s + 1} . \tag{20}$$

The average service rate of user $i$ is given as

$$\mu_i(\boldsymbol{q}) = \alpha_l \cdot \sum_{\boldsymbol{c} \in \boldsymbol{\mathcal{A}}} \left( \prod_{j \in \mathcal{I}} p_j^{c_j} \right) \xi_i(\boldsymbol{q}, \boldsymbol{c}) \cdot R_i^{c_i}, \ \forall i \in \mathcal{I}_l. \tag{21}$$

In each cluster, the dispatching policy of users can be calculated by solving an equation similar to (15), where the only difference is that the $\mu_i(\boldsymbol{q})$ in (15) is now given by (21).

### *B. Performance evaluation*

**Four-user heterogeneous scenario:** In this scenario the cell has two clusters ($\mathcal{I}_1$ and $\mathcal{I}_2$), and each cluster has two users ($U_1, U_2 \in \mathcal{I}_1, U_3, U_4 \in \mathcal{I}_2$). Fig. 13 (a) depicts the trade-off between performance and additional power expenditure. The maximal delay performance gain under traffic spreading is up to 22%, and up to 80% of the maximal gain is achieved at only 35% of the maximal re-routing cost. The split of average re-routing rates among users is shown in Fig. 13 (b). We can observe that users with worse channels ($U_2$ and $U_4$) also re-route files for users with better channels ($U_1$ and $U_3$).

**Users randomly distributed in a cell:** In this scenario we consider instances with five users, which are distributed uniformly in a cell at distances ranging from 50 to 200m to the BS. In each instance, the two users with the best average channel states act as Cluster Headers (CHs). CHs periodically broadcast their members to neighbouring users. A none-CH user chooses to join a cluster if it can communicate with all the current members of that cluster, augmented with tie-breaking rule. We evaluate 50 random instances and depict the average file transfer delay (as well as the $95^{th}$ and $5^{th}$ percentiles) in Fig. 13 (c). It is observed that the average delay under no re-routing can be improved by up to $40\%$ through the proposed traffic spreading. Moreover, with traffic spreading, the $95^{th}$ percentile of the delay under no re-routing is improved by up to 50%, while up to 73% of the maximal gain can be achieved by at only $18\%$ of the maximal re-routing cost.

## IX. Conclusion

In this paper, we presented a user-initiated traffic spreading approach, that is transparent to the BS, to improve the downlink delay performance in small cells. We formulated the problem of choosing the optimal dispatching policy as a Markov decision process and studied its properties in a two-user scenario. We also proposed a heuristic algorithm for multi-user scenarios. Our simulation results showed that the proposed approach can improve the delay performance greatly and the bulk of the performance can be achieved with a small increase in power expenditure. Moreover, even in the future when queue-aware scheduling policies are implemented at the BS, our proposed approach can still complement them and improve user performance.

## Appendix

### *A. Proof of Lemma 1*

To prove Lemma 1, we first derive the service rate $\boldsymbol{\mu}(\boldsymbol{q})$ under the delay-optimal scheduling policy (i.e. the Longest Connected Queue (LCQ) [30]). At any time $t$, a LCQ scheduler randomly chooses a queue $i$ to serve if it satisfies: *i)* Queue $i$ has the largest amount of work, where the work means the queue length divided by the service rate of the queue; *ii)* The achievable instantaneous service rate to user $i$ is positive at time $t$. Therefore, we can easily get the expression of $\boldsymbol{\mu}(\boldsymbol{q})$ under the LCQ scheduling policy:

$$\mu_i(\boldsymbol{q}) = \begin{cases} p_i, & q_i < q_j \\ (p_i + p_j - p_i p_j)/2, & q_i = q_j \\ p_i(1 - p_j), & q_i > q_j \end{cases} \tag{22}$$

where $i, j \in \{1,2\}$ and $i \neq j$. Then we make the following notations:

$$\mu_1' \equiv p_1, \quad \mu_1'' \equiv p_1(1-p_2) \tag{23}$$

$$\mu_2' \equiv p_2(1-p_1), \quad \mu_2'' \equiv p_2 \tag{24}$$

According to (22)-(24), we know

$$\mu_1' \geq \mu_1'', \quad \mu_2' \leq \mu_2'', \quad \mu_1' + \mu_2' = \mu_1'' + \mu_2'' \tag{25}$$

And when $p_1 = p_2$, we have

$$\mu_1' = \mu_2'', \quad \mu_1'' = \mu_2' \tag{26}$$

The uniform version of the continuous problem is present in section IV, with uniform rate $\varphi$ and transition probabilities shown in (8)-(10). We denote $w = w_j^i, \phi = \phi_j^i, \eta = \eta_j^i, \forall i, j \in \mathcal{I}$, then Bellman's equation takes the form

$$\varphi J_{k+1}(\boldsymbol{q}) = \varphi \frac{|\,\boldsymbol{q}\,|}{|\,\boldsymbol{\lambda}\,|} + \sum_{i \in \mathcal{I}} \mu_i(\boldsymbol{q}) J_k\big([D_i \boldsymbol{q}]^+\big) \tag{27}$$

$$+ \min_{\boldsymbol{\sigma}(\boldsymbol{q}) \in \mathcal{A}} \sum_{j \in \mathcal{I}} \sum_{i \in \mathcal{I}} \lambda_j \sigma_j^i(\boldsymbol{q}) \big[z_j^i \cdot (\eta + w \cdot \phi) + J_k\big(A_i \boldsymbol{q}\big)\big]$$

*Poof of Lemma 1:* We prove this lemma using induction. First we make the following definitions:

$$\begin{aligned} f_1(\boldsymbol{q}) \equiv & \mu_1(\boldsymbol{q}+\boldsymbol{e}_1) J_k(\boldsymbol{q}) + \mu_2(\boldsymbol{q}+\boldsymbol{e}_1) J_k([\boldsymbol{q}+\boldsymbol{e}_1-\boldsymbol{e}_2]^+) \\ & - \mu_1(\boldsymbol{q}+\boldsymbol{e}_2) J_k([\boldsymbol{q}-\boldsymbol{e}_1+\boldsymbol{e}_2]^+) - \mu_2(\boldsymbol{q}+\boldsymbol{e}_2) J_k(\boldsymbol{q}) \\ f_2(\boldsymbol{q}) \equiv & \lambda_1 \big\{ \min\big[J_k(\boldsymbol{q}+2\boldsymbol{e}_1), \eta + w\phi + J_k(\boldsymbol{q}+\boldsymbol{e}_1+\boldsymbol{e}_2)\big] \\ & - \min\big[J_k(\boldsymbol{q}+\boldsymbol{e}_1+\boldsymbol{e}_2), \eta + w\phi + J_k(\boldsymbol{q}+2\boldsymbol{e}_2)\big]\big\} \\ f_3(\boldsymbol{q}) \equiv & \lambda_2 \big\{ \min\big[\eta + w\phi + J_k(\boldsymbol{q}+2\boldsymbol{e}_1), J_k(\boldsymbol{q}+\boldsymbol{e}_1+\boldsymbol{e}_2)\big] \\ & - \min\big[\eta + w\phi + J_k(\boldsymbol{q}+\boldsymbol{e}_1+\boldsymbol{e}_2), J_k(\boldsymbol{q}+2\boldsymbol{e}_2)\big]\big\} \end{aligned}$$

*Induction basis:* Since $J_0(\boldsymbol{q}) = 0, \forall \boldsymbol{q} \in \mathbb{Z}^2$, we know $\Delta_0(\boldsymbol{q})$ is monotonically non-decreasing in $q_1$.

*Induction step:* Assume $\Delta_k(\boldsymbol{q})$ is monotonically non-decreasing in $q_1$ for each fixed $q_2$. By substituting (27) into (19), we have

$$\begin{aligned} \varphi \Delta_{k+1}(\boldsymbol{q}) = & \varphi\big(J_{k+1}(\boldsymbol{q}+\boldsymbol{e}_1) - J_{k+1}(\boldsymbol{q}+\boldsymbol{e}_2)\big) \\ = & f_1(\boldsymbol{q}) + f_2(\boldsymbol{q}) + f_3(\boldsymbol{q}) \end{aligned} \tag{28}$$

In the following we show that all $f_1(\boldsymbol{q}), f_2(\boldsymbol{q})$ and $f_3(\boldsymbol{q})$ are monotonically non-decreasing in $q_1$ for each fixed $q_2$.

$f_1(\boldsymbol{q})$: We have to consider different queue states around the diagonal, as shown in Fig 14. According to different queue state, we prove this part by five different cases.

*Case 1:* If $q_1 + 1 < q_2$, according to (22)-(24) we know $\mu_1(\boldsymbol{q}+\boldsymbol{e}_1) = \mu_1(\boldsymbol{q}+\boldsymbol{e}_2) = \mu_1''$ and $\mu_2(\boldsymbol{q}+\boldsymbol{e}_1) = \mu_2(\boldsymbol{q}+\boldsymbol{e}_2) = \mu_2''$. Thus $f_2(\boldsymbol{q})$ can be written as the following:

$$\begin{aligned} f_1(\boldsymbol{q}) \equiv & \mu_1''[J_k(\boldsymbol{q}) - J_k([\boldsymbol{q}-\boldsymbol{e}_1+\boldsymbol{e}_2]^+)] \\ & + \mu_2''[J_k([\boldsymbol{q}+\boldsymbol{e}_1-\boldsymbol{e}_2]^+ - J_k(\boldsymbol{q})] \\ = & \mu_1''\Delta_k([\boldsymbol{q}-\boldsymbol{e}_1]^+) + \mu_2''\Delta_k([\boldsymbol{q}-\boldsymbol{e}_2]^+) \end{aligned} \tag{29}$$

which is non-decreasing in $q_1$ from the induction.

*Case 2:* If $q_1 + 1 = q_2$, we have $\mu_1(\boldsymbol{q}+\boldsymbol{e}_1) = \mu_1', \mu_1(\boldsymbol{q}+\boldsymbol{e}_2) = \mu_1''$ and $\mu_2(\boldsymbol{q}+\boldsymbol{e}_1) = \mu_2', \mu_2(\boldsymbol{q}+\boldsymbol{e}_2) = \mu_2''$ according to (22)-(24). Furthermore, we get $J_k(\boldsymbol{q}) = J_k(\boldsymbol{q}+\boldsymbol{e}_1-\boldsymbol{e}_2)$

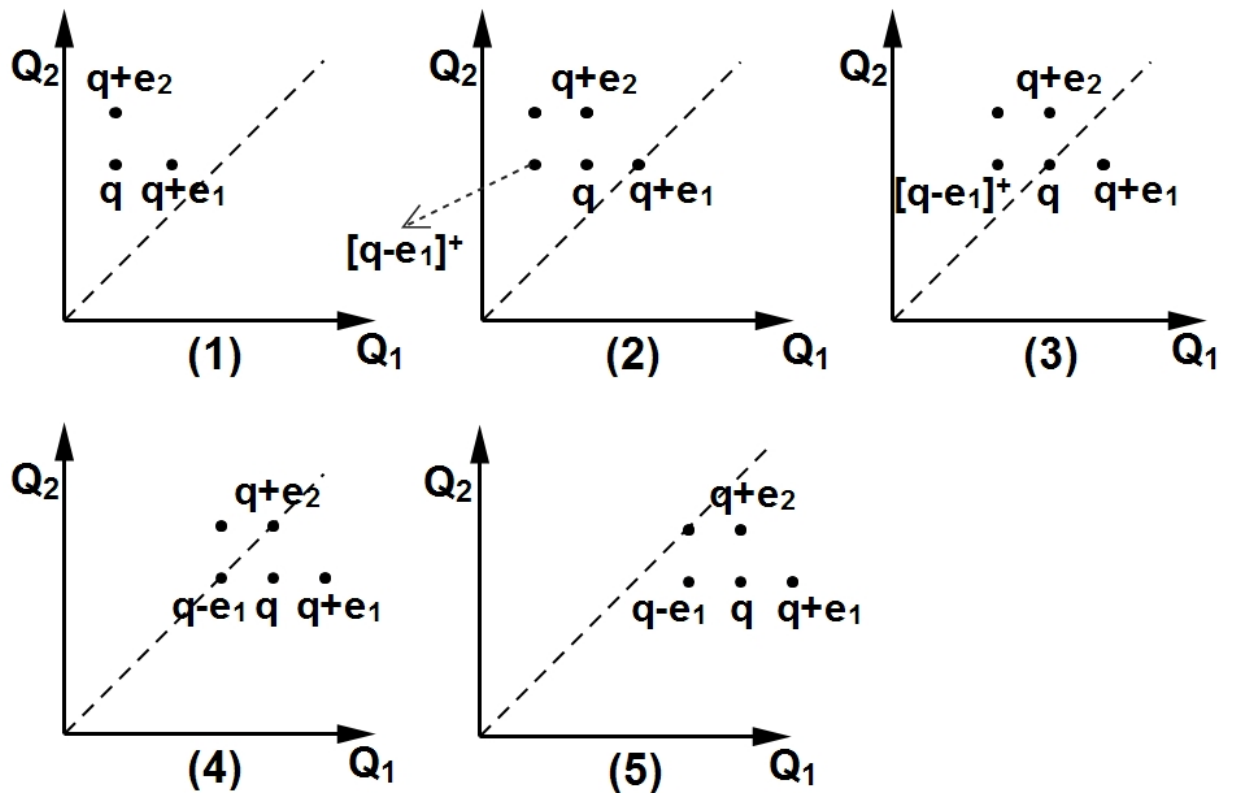

Fig. 14: Different queue states $\boldsymbol{q} \in \mathbb{Z}_+^2$ around the diagonal.

since $q_1 + 1 = q_2$. Thus $f_2(\boldsymbol{q})$ can be written as:

$$\begin{aligned} f_1(\boldsymbol{q}) \equiv & \mu_1''[J_k(\boldsymbol{q}) - J_k([\boldsymbol{q}-\boldsymbol{e}_1+\boldsymbol{e}_2]^+)] \\ & + \mu_2''[J_k(\boldsymbol{q}+\boldsymbol{e}_1-\boldsymbol{e}_2) - J_k(\boldsymbol{q})] \\ = & \mu_1''\Delta_k([\boldsymbol{q}-\boldsymbol{e}_1]^+) + \mu_2''\Delta_k(\boldsymbol{q}-\boldsymbol{e}_2) \end{aligned} \tag{30}$$

which is non-decreasing in $q_1$ from the induction.

*Case 3:* If $q_1 = q_2$, we have $J_k([\boldsymbol{q}-\boldsymbol{e}_2]^+) = J_k([\boldsymbol{q}-\boldsymbol{e}_1]^+)$.Besides, we know $\mu_1(\boldsymbol{q}) = \mu_2(\boldsymbol{q}) = (\mu_1' + \mu_2')/2$. Thus $f_2(\boldsymbol{q}) - f_2([\boldsymbol{q}-\boldsymbol{e}_1]^+)$ is

$$\begin{aligned} f_1(\boldsymbol{q}) - f_1([\boldsymbol{q}-\boldsymbol{e}_1]^+) = & \ \mu_1' J_k(\boldsymbol{q}) + \mu_2' J_k([\boldsymbol{q}+\boldsymbol{e}_1-\boldsymbol{e}_2]^+) \\ & - \mu_1'' J_k([\boldsymbol{q}-\boldsymbol{e}_1+\boldsymbol{e}_2]^+) - \mu_2'' J_k(\boldsymbol{q}) \\ & - \frac{\mu_1'' + \mu_2''}{2} \Big\{ J_k([\boldsymbol{q}-\boldsymbol{e}_1]^+) + J_k([\boldsymbol{q}-\boldsymbol{e}_2]^+) \Big\} \\ & + \mu_1'' J_k([\boldsymbol{q}-2\boldsymbol{e}_1+\boldsymbol{e}_2]^+) + \mu_2'' J_k([\boldsymbol{q}-\boldsymbol{e}_1]^+) \\ = & - \big\{\mu_1'' J_k([\boldsymbol{q}-\boldsymbol{e}_1]^+) - \mu_1'' J_k([\boldsymbol{q}-2\boldsymbol{e}_1+\boldsymbol{e}_2]^+)\big\} \\ = & - \Delta_k([\boldsymbol{q}-2\boldsymbol{e}_1]^+) \geq -\Delta_k(\boldsymbol{q}) = 0 \end{aligned} \tag{31}$$

*Case 4:* The proof under this case is similar to *Case 2*.

*Case 5:* The proof under this case is similar to *Case 1*.

$f_2(\boldsymbol{q})$: $f_2(\boldsymbol{q})$ can be written as follows:

$$\begin{aligned} f_2(\boldsymbol{q}) = & \lambda_1 \big\{ J_k(\boldsymbol{q}+\boldsymbol{e}_1+\boldsymbol{e}_2) - J_k(\boldsymbol{q}+\boldsymbol{e}_1+\boldsymbol{e}_2) \\ & + \min\big[J_k(\boldsymbol{q}+2\boldsymbol{e}_1) - J_k(\boldsymbol{q}+\boldsymbol{e}_1+\boldsymbol{e}_2), \eta + w\phi\big] \\ & - \min\big[0, \eta + w\phi + J_k(\boldsymbol{q}+2\boldsymbol{e}_2) - J_k(\boldsymbol{q}+\boldsymbol{e}_1+\boldsymbol{e}_2)\big]\big\} \\ = & \lambda_1 \big\{ \min\big[\Delta_k(\boldsymbol{q}+\boldsymbol{e}_1), \eta + w\phi\big] \\ & + \max\big[0, \Delta_k(\boldsymbol{q}+\boldsymbol{e}_2) - \eta - w\phi\big]\big\} \\ = & \lambda_1 \big\{ \eta + w\phi + \min\big[0, \Delta_k(\boldsymbol{q}+\boldsymbol{e}_1) - \eta - w\phi\big] \\ & + \max\big[0, \Delta_k(\boldsymbol{q}+\boldsymbol{e}_2) - \eta - w\phi\big]\big\} \end{aligned} \tag{32}$$

where $\Delta_k(\boldsymbol{q}+\boldsymbol{e}_1)$ and $\Delta_k(\boldsymbol{q}+\boldsymbol{e}_2)$ are monotonically non-decreasing in $q_1$ for each fixed $q_2$ from induction, resulting in $\min\big[0, \Delta_k(\boldsymbol{q}+\boldsymbol{e}_1) - \eta - w\phi\big]$ and $\Delta_k(\boldsymbol{q}+\boldsymbol{e}_2) - \eta - w\phi\big]$ are non-decreasing in $q_1$. Therefore, $f_2(\boldsymbol{q})$ is non-decreasing in $q_1$ for each fixed $q_2$.

$f_3(\boldsymbol{q})$: Similarly, $f_3(\boldsymbol{q})$ is non-decreasing in $q_1$ for each fixed $q_2$, by written as follows:

$$\begin{aligned} f_3(\boldsymbol{q}) = & \lambda_2 \big\{ -w\phi + \min\big[0, \Delta_k(\boldsymbol{q}+\boldsymbol{e}_1) + \eta + w\phi\big] \\ & + \max\big[0, \Delta_k(\boldsymbol{q}+\boldsymbol{e}_2) + \eta + w\phi\big]\big\} \end{aligned}$$

Therefore, this lemma is proved since all $f_1(\boldsymbol{q}), f_2(\boldsymbol{q})$ and $f_3(\boldsymbol{q})$ are monotonically non-decreasing in $q_1$ for each fixed $q_2$. ■